\documentclass[aps,prx,twocolumn,superscriptaddress,final]{revtex4-2}
\usepackage{amsmath}
\usepackage{amsfonts}
\usepackage{amssymb}
\usepackage{tikz}
\usetikzlibrary{positioning, arrows.meta, calc}
\usepackage{graphicx}
\usepackage{xcolor}
\usepackage[colorlinks=True,citecolor=mydarkblue,linkcolor=mydarkblue,urlcolor=mygreen]{hyperref}
\usepackage{hypcap}
\usepackage{braket}
\usepackage{bm}
\usepackage{dsfont}
\usepackage[export]{adjustbox} 
\usepackage{verbatim}

\usepackage{csquotes}
\usepackage[backgroundcolor=white,bordercolor=red,linecolor=red,textcolor=red,textsize=footnotesize]{todonotes}

\usepackage[capitalise]{cleveref} 
\definecolor{mydarkblue}{RGB}{0,0,130}
\definecolor{myblue}{RGB}{0,0,200}
\definecolor{myorange}{RGB}{130,50,0}
\definecolor{mygreen}{RGB}{0,130,0}

\newcommand{\eye}{\mathds{1}} 
\newcommand{\SU}[1]{$\mathrm{SU}(#1)$} 
\renewcommand{\O}[1]{$\mathrm{O}(#1)$} 

\Crefname{section}{Sec.}{Secs.}

\begin{document}
\title{Post-measurement Quantum Monte Carlo}
\author{Kriti Baweja}
\affiliation{Institute of Physics, University of Bonn, Nußallee 12, 53115 Bonn, Germany}
\author{David J. Luitz}
\affiliation{Institute of Physics, University of Bonn, Nußallee 12, 53115 Bonn, Germany}
\author{Samuel J. Garratt}
\affiliation{Department of Physics, University of California, Berkeley, CA 94720, USA}

\begin{abstract}
We show how the effects of large numbers of measurements on many-body quantum ground and thermal states can be studied using Quantum Monte Carlo (QMC). Density matrices generated by measurement in this setting feature products of many local nonunitary operators, and by expanding these density matrices as sums over operator strings we arrive at a generalized stochastic series expansion (SSE). Our `post-measurement SSE' is based on importance sampling of operator strings contributing to a measured thermal density matrix. We demonstrate our algorithm by probing the effects of measurements on the spin-$1/2$ Heisenberg antiferromagnet on the square lattice. Thermal states of this system have \SU{2} symmetry, and at first we preserve this symmetry by measuring \SU{2} symmetric observables. We identify classes of post-measurement states for which correlations can be calculated efficiently, as well as states for which \SU{2} symmetric measurements generate a QMC sign problem when working in any site-local basis. For the first class, we show how deterministic loop updates can be leveraged. Using our algorithm we demonstrate the creation of long-range Bell pairs and symmetry-protected topological order, as well as the measurement-induced enhancement of antiferromagnetic correlations. We then study the effects of measuring the system in a basis where the standard (unmeasured) SSE is sign-free: For measurement schemes with this property, we can calculate correlations in all post-measurement states without a sign problem. The method developed in this work opens the door to scalable experimental probes of measurement-induced collective phenomena, which require numerical estimates for the effects of measurements.
\end{abstract}

\maketitle

\section{Introduction}

Measuring part of a many-body quantum system can generate new kinds of correlations between the unmeasured degrees of freedom. Even weakly entangled states, having exact tensor network descriptions and finite-range correlations, can become extremely complex when we perform large numbers of measurements. This phenomenon is the basis of measurement-based quantum computation \cite{Briegel_2009}, as well as of efforts to demonstrate quantum advantage in random circuit sampling \cite{Arute_2019}. The complexity of this sampling task is closely related \cite{napp2022efficient} to the existence of measurement-induced entanglement transitions in monitored quantum dynamics \cite{skinner2019measurement,li2018quantum,fisher2023random}. Local measurements additionally provide a direct route from trivial product states to topological order and long-range entanglement \cite{raussendorf2005long,aguado2008creation,piroli2021quantum,tantivasadakarn2024long,lu2022measurement,zhu2023nishimori,lee2022decoding}. 

These results raise the question of how measurements restructure ground and thermal states of physical quantum systems. At high temperatures, thermal states are exactly separable and the effects of local measurements are short-ranged \cite{yin2023polynomial,bakshi2024high}. At low energies, on the other hand, local degrees of freedom develop nontrivial correlations, and the nonlocal effects of measurements can be dramatic. In particular, measurements performed in different locations can conspire with one another to qualitatively alter correlations and entanglement over large length scales \cite{garratt2023measurements,lee2023quantum,murciano2023measurement,weinstein2023nonlocality,yang2023entanglement,sun2023new,su2024higher,sala2024quantum,hoshino2024entanglement,patil2024highly}. These phenomena can be analyzed via the quantum-classical mapping: measurements appear as defects, introduced at a fixed `imaginary time', in the Euclidean spacetime path integrals which represent the initial thermal density matrices \cite{rajabpour2015post,garratt2023measurements}.

\begin{figure}[ht]
\includegraphics[width=\columnwidth]{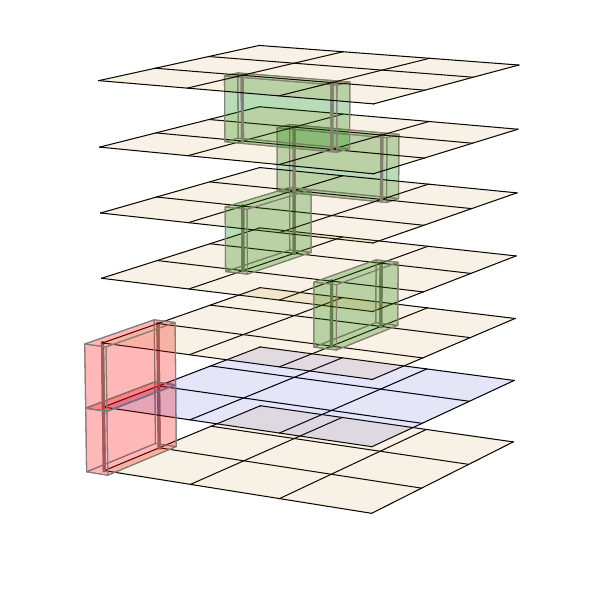}
\caption{Operator string in the post-measurement stochastic series expansion for a system in two spatial dimensions. Operators from the expansion of the Hamiltonian are shown as green boxes, while operators from the (two-site) measurements are shown in red. The different planes indicate the state of the system at different ``imaginary time slices'', with implied periodicity between the top and the bottom slice.}
\label{fig:proj_decomp}
\end{figure}

This description indicates a general connection between the effects of measurements on low-energy many-body states and the statistical mechanics of surfaces. It is therefore natural to expect a wide variety of new phenomena beyond one spatial dimension, but their investigation demands new numerical techniques. In this work we develop a new quantum Monte Carlo (QMC) algorithm allowing for the calculation of `post-measurement' correlations in ground and thermal quantum states. Our algorithm is based on a modification of the stochastic series expansion (SSE) \cite{sandvik_computational_2010}. We consider expansions of measured many-body density matrices as sums over products (strings) of local operators and, by designing efficient methods for sampling these operator strings, we show how to efficiently calculate post-measurement expectation values. An illustration of the contributing operator strings is shown in Fig.~\ref{fig:proj_decomp}.

Previous numerical methods to study the effects of measurements on correlated low-energy quantum states have been based on matrix product states \cite{lin2023probingsign,garratt2023measurements,murciano2023measurement,yang2023entanglement,sun2023new,sala2024quantum,hoshino2024entanglement} or on Gaussian states \cite{weinstein2023nonlocality,murciano2023measurement,sala2024quantum,cheng2024universal}, and so have been limited to one spatial dimension or to non-interacting systems, respectively. The post-measurement SSE significantly extends the scope of these investigations, giving access to large classes of interacting quantum systems in arbitrary spatial dimensions. This extension is especially important for quantum critical systems; there the effects of measurements can be striking, but the Mermin-Wagner theorem places strong restrictions on the emergence of critical behavior in one spatial dimension \cite{sachdev2011}. 

An additional motivation for this work comes from the role of numerics in scalable experiments on measurement-induced phenomena \cite{gullans2020scalable,li2021robust,noel2022measurement,li2023cross,garratt2023measurements,google2023measurement,garratt2023probing,mcginley2024postselection,tikhanovskaya2024universality,kamakari2024experimental}. This role has its origin in the `post-selection problem': The probability to prepare any specific post-measurement state is, in general, exponentially small in the number of measurements, so determining the expectation value of an observable in such a state requires exponential time. Cross-correlations between the results of experiments and simulations can nevertheless provide rigorous bounds on properties of post-measurement quantum states, allowing us to efficiently diagnose the effects of our measurements \cite{garratt2023probing,mcginley2024postselection}. 

The post-measurement SSE therefore enables experiments on measurement-induced collective phenomena across a wide variety of platforms for quantum simulation. The basic experimental requirements are the ability to prepare low-energy many-body states, as well as access to local measurements. These requirements have been satisfied for some time in Rydberg tweezer arrays \cite{browaeys2020many}, ultracold quantum gases in optical lattices \cite{gross2017quantum}, ion traps \cite{monroe2021programmable}, and systems of superconducting qubits \cite{kjaergaard2020superconducting}. The measurements which create the state of interest can be performed simultaneously with the measurements that are used to probe it, so  `mid-circuit' measurements \cite{minev2019catch,iqbal2024topological,bluvstein2024logical} are not required. 

Our numerical method can be applied to study the effects of measurements on ground and thermal states provided there exists a basis for which the standard (unmeasured) SSE is free of a QMC sign problem. Measurements of operators that are diagonal in that sign-free basis can then be introduced to construct post-measurement SSEs that are themselves sign-free for all possible measurement outcomes. However, measurements of operators that are \emph{not} diagonal in the sign-free basis generically lead to post-measurement SSEs that do have sign problems. 

Here our focus is on the effects of measurements on the spin-$1/2$ antiferromagnetic (AFM) Heisenberg model on the square lattice. This model has \SU{2} symmetry and, in the case of isotropic nearest-neighbor interactions, it exhibits long-range order at vanishing temperatures. By varying the interactions to favor dimerization, the ground state can be driven through a quantum phase transition to a disordered phase. The transition between these phases is widely expected to be described by the three-dimensional \O{3} universality class \cite{chakravarty1989two}. 

Through the majority of this work, the measurements we study preserve the \SU{2} symmetry of the state. By measuring the total spin on nearby pairs of spin-$1/2$ degrees of freedom we generate either local singlet or triplet degrees of freedom. While a singlet outcome necessarily disentangles the measured spins from the rest of the system, a triplet outcome generates an effective spin-$1$ degree of freedom, and these local three-level systems can remain entangled with their surroundings.

We choose this setup to illustrate a variety of features of the post-measurement SSE. For generic measurement outcomes, and when working with a basis of product states, the post-measurement SSE features a sign problem; the sign problem which we generate has a similar origin those familiar in the study of frustrated antiferromagnets \cite{sandvik_computational_2010}. The presence of a sign problem in the post-measurement SSE is to be contrasted with its absence in the standard (thermal) SSE for our model. For the subset of post-measurement states whose SSEs do not suffer from a sign problem, we show that the operator strings contributing to the density matrix can be sampled using loop updates, thereby allowing for efficient calculations of observable expectation values. We use these states (which, in experiment, would arise from post-selection) to explore the implications of the \SU{2} symmetry for post-measurement correlations. 

First, we study the generation of long-range singlet states and symmetry-protected topological (SPT) \cite{senthil2015symmetry} states using local measurements. We show also that the symmetry of the system allows for a straightforward calculation of mixed-state entanglement measures using QMC. Following this, we show that the interplay between correlations in the initial thermal density matrix and (non-overlapping) local measurements can enhance long-range correlations. Using these measurements we probe an \O{3} quantum critical point, and provide evidence that in certain post-selected states the correlations decay parametrically slower than in the ground state. This behavior is expected for a quantum state whose correlations realize the extraordinary-log boundary universality class of the three-dimensional \O{3} model \cite{metlitski2022boundary,lee2023quantum}.  

Finally, we study the measurement of standard Pauli operators. In this case, when all measurements are in the same basis, for all possible measurement outcomes the post-measurement SSE has no sign problem. Because of this we can efficiently calculate averaged properties of the ensemble of post-measurement states. At the \O{3} quantum critical point, such measurements are expected to behave as a relevant perturbation \cite{garratt2023measurements}; we verify numerically that the critical correlations are destroyed when we measure the spins on one of the two sublattices.

This work is organized as follows. In Sec.~\ref{sec:background} we provide background on the analysis of measurement-induced phenomena, the model studied in this work, and the standard SSE. The post-measurement SSE is introduced in Sec.~\ref{sec:algorithms}; there we discuss how the effects of both projective and weak measurements can be captured within a QMC calculation. In Sec~\ref{sec:singlet} we demonstrate the creation of long-range singlet and SPT states. In Sec.~\ref{sec:triplet} we study the measurement-induced enhancement of antiferromagnetic correlations in post-selected states. Then, in Sec.~\ref{sec:ensemble}, we study the full ensemble of post-measurement states generated by measurements of local spin operators. We discuss our results as well as open questions in Sec.~\ref{sec:discussion}.

\section{Background}\label{sec:background}
In this section we provide the background necessary for the implementation of the algorithm. In Sec.~\ref{sec:ovmeas} we describe the effects of measurements on ground and thermal quantum states, outlining how these effects can be understood by considering Euclidean time path integrals. In Sec.~\ref{sec:modelandsymm} we describe the model and the measurements studied in this work. In Sec.~\ref{sec:intro_SSE} we review the standard SSE, which allows for the calculation of equal-time correlations in thermal states.

\subsection{Post-measurement states}\label{sec:ovmeas}

Here we describe basic properties of post-measurement states, the kinds of physical quantities which can probe the effects of measurements, and how the quantum-classical mapping can be leveraged in this setting. 

The setup is as follows. Suppose we have a set of $N$ qubits labelled $j=0,\ldots,(N-1)$, and that the system is initialized in a thermal state $\rho = \frac{1}{Z} e^{-\beta H}$ with respect to a Hamiltonian $H$. We then consider the effect of measuring a set of observables supported within a subregion $B$, for example one-body observables on sites $j \in B$, or two-body observables on pairs of sites $(jk)$, with each of $j,k \in B$; we will also write this as $(jk) \in B$. We denote by $A$ the complement of the measured region $B$. In each case we denote the set of outcomes by $\bm{s}$, i.e. $\bm{s}=\{s_j|j \in B\}$ for one-body measurements, or $\bm{s}=\{s_{(jk)}|(jk) \in B\}$ for two-body measurements. For projective measurements we can associate the full set of outcomes with a product of projectors, such as $P^{\bm{s}}=\bigotimes_{j \in B}P^{s_j}_j$ in the one-body case, where $P^{s_j}_j$ is a projection operator on site $j$ which depends on the outcome $s_j$. The probability $p^{\bm{s}}$ to observe the set of outcomes $\bm{s}$ is 
\begin{align}
    p^{\bm{s}} = \text{Tr}[\rho P^{\bm{s}}].
\end{align}
The post-measurement density matrix which we create in this case is
\begin{align}
    \rho^{\bm{s}} = \frac{1}{p^{\bm{s}}}P^{\bm{s}} \rho P^{\bm{s}}.
    \label{eq:rhos}
\end{align}

We are interested in correlations and entanglement in post-measurement states $\rho^{\bm{s}}$. For example, the post-measurement correlation function $\braket{X_j X_k}^{\bm{s}}$ is
\begin{align}
    \braket{X_j X_k}^{\bm{s}} = \text{Tr}[X_j X_k \rho^{\bm{s}}] = \text{Tr}[X_j X_k \rho^{\bm{s}}_{jk}],
\end{align}
where $\rho^{\bm{s}}_{jk}$ is the reduced density matrix for qubits $j$ and $k$ obtained by tracing out all other qubits. Post-measurement correlations can be nontrivial because the effects of measurements are nonlocal. However, it is important to note that these nonlocal effects can only arise if we keep track of the outcomes. If we do not keep track of the outcomes, e.g. by averaging a quantity $ \braket{X_j X_k}^{\bm{s}}$ that is \emph{linear} in $\rho^{\bm{s}}$ over outcomes $\bm{s}$, we obscure the effects of our measurements. This is because measuring a local observable and then averaging over the different outcomes has an equivalent effect on the density matrix as a local quantum channel $\Phi[\rho]=\sum_{s} K^s \rho [K^s]^{\dag}$, where $\sum_{s} [K^s]^{\dag} K^s = \eye$ for normalization, and a local quantum channel has strictly local effects. For example, if the Kraus operators $K^s$ have support only in a region $B$, then $\text{Tr}_B\Phi[\rho]=\rho$. Above, the Kraus operators are the projectors $P^{\bm{s}}$, so if we do not measure the sites $j$ and $k$,
\begin{align}
    \sum_{\bm{s}} p^{\bm{s}}\rho^{\bm{s}}_{jk} = \rho_{jk},
\end{align}
where $\rho_{jk}$ is the unmeasured reduced density matrix for sites $j$ and $k$. In order to study measurement-induced collective phenomena, there are two possibilities: Average quantities nonlinear in $\rho^{\bm{s}}$, such as $\sum_{\bm{s}}p^{\bm{s}} \braket{X_j}^{\bm{s}}\braket{X_k}^{\bm{s}}$, or calculate correlations in individual post-measurement states $\rho^{\bm{s}}$. We analyze both cases in this work.

We will also consider the effects of `weak' measurements. Physically, weak measurements arise from coupling auxiliary degrees of freedom to the system, and then measuring the auxiliary degrees of freedom. A weak measurement of an observable on site $j$, having outcome $s_j$, corresponds to Kraus operator $W^{s_j}_{\mu,j}$. The `strength' of the measurement is denoted $\mu$, and for $\mu \to \infty$ our weak measurement becomes projective, $W^{s_j}_{\mu,j} \to P^{s_j}_j$. For weak measurements on all $j \in B$, the full measurement operator is $W^{\bm{s}}_{\mu} = \bigotimes_{j \in B} W^{s_j}_{\mu,j}$. In this case, the probability $p^{\bm{s}}_{\mu}$ for the set of outcomes $\bm{s}$, and the post-measurement density matrix $\rho^{\bm{s}}_{\mu}$, are given by
\begin{align}
    p^{\bm{s}}_{\mu} &= \text{Tr}[W^{\bm{s}}_{\mu}\rho (W^{\bm{s}}_{\mu})^{\dag}],   \quad \rho^{\bm{s}}_{\mu} = \frac{1}{p^{\bm{s}}_{\mu}} W^{\bm{s}}_{\mu} \rho (W^{\bm{s}}_{\mu})^{\dag}.
\end{align}
The probability conservation condition is $\sum_{\bm{s}}(W^{\bm{s}}_{\mu})^{\dag}W^{\bm{s}}_{\mu} = \eye$, and below we will restrict to Hermitian $W^{\bm{s}}_{\mu}$. In the projective limit $\mu \to \infty$ we omit the $\mu$ subscript, e.g. $\rho^{\bm{s}}_{\infty}=\rho^{\bm{s}}$.

The effects of measurements on ground and thermal states can be understood via the quantum-classical mapping \cite{rajabpour2015post,garratt2023measurements}. On writing the thermal density matrix as a Euclidean spacetime path integral, spatial (equal-time) correlations in this state become correlations within a $d$-dimensional surface of the $(d+1)$-dimensional spacetime. We refer to this surface as `$\tau=0$', where the imaginary (or Euclidean) time $\tau \in [0,\beta]$ runs around a thermal circle of circumference $\beta$. When calculating post-measurement correlation functions using this formalism, measurements appear as defects in the $\tau=0$ surface. This follows simply from the fact that in the post-measurement density matrix $\rho^{\bm{s}} \sim P^{\bm{s}}e^{-\beta H} P^{\bm{s}}$ the projectors $P^{\bm{s}}$ appear on either side of the imaginary-time evolution operator $e^{-\beta H}$, rather than `in the middle of' this evolution. 

\subsection{Model}\label{sec:modelandsymm}

To demonstrate our algorithm we will study the effects of measurements on thermal states of the spin-$1/2$ Heisenberg AFM on the square lattice. The Hamiltonian is
\begin{align}
	H = \sum_{(j, k) \in \text{NN}} J_{jk}\,\,\mathbf{S}_j \cdot \mathbf{S}_k
 \label{eqn:AFM_ham}
\end{align}
where $j=0, \ldots, (N-1)$ labels sites and NN is the set of nearest neighbors. The operators $\bm{S}_j=(S^x_j,S^y_j,S^z_j)$ satisfy the standard commuation relations $[S^a_j,S^b_k]=i\delta_{jk}\epsilon^{abc}S^c_j$, and the initial unmeasured density matrix is $\rho = e^{-\beta H}/\text{Tr}e^{-\beta H}$. We use periodic boundary conditions along both directions, with an even number of lattice sites along each to prevent frustration. In systems having length $N_x$ in the $x$ direction and $N_y$ in the y direction, our convention is that the coordinates of site $j$ are $(x_j,y_j)$ with $x_j = j \bmod{N_x}$ and $y_j = (j-x_j)/ N_x$.  The couplings $J_{jk} \geq 0$ are unfrustrated, so the ground state of $H$ is a singlet under global \SU{2} transformations \cite{auerbach2012interacting}.

\begin{figure}
    \centering
 \begin{tikzpicture}[scale=1.5]
    \foreach \x in {0,1,2,3} {
        \draw[very thick] (0,\x) -- (3,\x);
        \draw[very thick] (\x,0) -- (\x,3);
    }

    \foreach \x in {0,1,2,3} {   
        \foreach \y in {0,2} {
          \draw[red!85!black,line width=1mm] (\x,\y) --  (\x,\y+1);
        }
    }

    \node[red!85!black, anchor=west] at (0,0.5)  {$J_2$};
    \node[anchor=west] at (0,1.5)  {$J_1$};

    \foreach \x in {0,1,2,3} {
        \foreach \y in {0,1,2,3} {
            \node[draw, thick, circle, minimum size=0.2mm, fill=white] at (\x,\y) {};
        }
    }
\end{tikzpicture}
    \caption{AFM Heisenberg model on a square lattice. Interactions $J_1 \bm{S}_j \cdot \bm{S}_k$ and $J_2 \bm{S}_j \cdot \bm{S}_k$ are indicated by black and red NN bonds, respectively. The ground state of the isotropic model $J_1=J_2$ has long-range Ne\'{e}l order, while for large $J_2/J_1$ it is a short-range correlated paramagnet.}
    \label{fig:lat_j1_j2}
\end{figure}
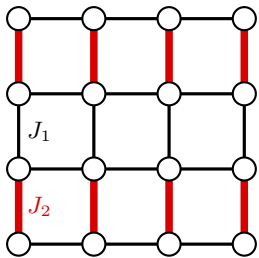

In the isotropic Heisenberg model $J_{jk}=J$ the system is long-range ordered at zero temperature ($\beta \to \infty$), but is disordered for finite $\beta$ as a consequence of the Mermin-Wagner theorem. We study this isotropic model as well as systems with interactions favoring dimerization. In general we set $J_{jk}=J_2$ for even NN pairs of sites $(jk)$, i.e. pairs $(jk)$ with $x_j=x_k$ and $y_k=y_j+1$ where $y_j=2n$, $n\in  \{0 \cdots N_y/2-1\}$, while we set $J_{jk}=J_1$ on all other bonds; this arrangement of interaction strength is illustrated in \cref{fig:lat_j1_j2}. The competition of the $J_1$ and $J_2$ interactions induces a quantum phase transition between an ordered and a dimerized phase when the ratio $g=J_2/J_1$ is exceeds $g_c \approx 1.91$ \cite{wenzel2009comprehensive}. The transition between these two phases is expected to belong to the \O{3} universality class.

To preserve the \SU{2} symmetry of the density matrix, we consider measurements of \SU{2} symmetric local operators. 
Specifically, we measure the magnitude of the composite spin on a pair of sites, i.e. $S_{(jk)}^2 = \left(\bm{S}_j + \bm{S}_k \right)^2$, with $(jk)$ nearest neighbors (NN) or next-nearest neighbor (NNN). 
Moreover, we restrict ourselves to measurements of non-overlapping pairs of sites, although it is important to note that this restriction is not necessary for a QMC scheme. The outcomes of our measurements are the different possible values of the total spin on the pair of sites $(jk)$, i.e. $s_{(jk)}=0$ for a singlet outcome or $s_{(jk)}=1$ for a triplet outcome, where $s_{(jk)}(s_{(jk)}+1)$ is the corresponding eigenvalue of $S_{(jk)}^2$. For a projective measurement with a singlet outcome the state of the measured pair of sites is uniquely determined, but for a triplet outcome it hosts a spin-$1$ degree of freedom. In that case, the measurement of $S_{(jk)}^2$ does not determine, e.g. the eigenvalue with respect to $S^z_j+S^z_k$. 

The projection operators corresponding to these outcomes are denoted $P^{s_{(jk)}}_{(jk)}$. For a singlet outcome
\begin{equation}
    P_{(jk)}^0 =\frac{1}{4}-\bm{S}_j \cdot \bm{S}_k,
    \label{eqn:proj_defn_singlet}
\end{equation}
and for a triplet outcome
\begin{equation}
    P_{(jk)}^1 =\frac{3}{4} +\bm{S}_j \cdot \bm{S}_k.
\label{eqn:proj_defn_triplet}
\end{equation}
If we measure the total spin on many pairs of sites $(jk)$, finding outcomes $\bm{s}=\{ s_{(jk)}|(jk) \in B\}$, where $B$ is the set of measured pairs, the full projection operator is 
\begin{align}
    P^{\bm{s}} = \bigotimes_{(jk) \in B} P^{s_{(jk)}}_{(jk)}.
\end{align}
This is the form of the measurement operator that we will use for our numerical calculations in Sec.~\ref{sec:algorithms} - ~\ref{sec:singlet}. The single-site measurements discussed in Sec.~\ref{sec:ovmeas} will be used to illustrate the process of sampling in Sec.~\ref{sec:intro_SSE} and will be investigated in detail in Sec.~\ref{sec:ensemble}.

For all $\beta$ the thermal density matrix has a weak symmetry $U\rho U^{-1}=\rho$ under global \SU{2} rotations $U$, and because of the \SU{2} symmetry of $\bm{S}_j\cdot \bm{S}_k$ this feature is shared with the post-measurement states generated by measurements of $S_{jk}^2$. Similarly, for $\beta \to \infty$ these states have strong \SU{2} symmetry, $\rho = U\rho=\rho U$, since they are total singlets.

\subsection{Stochastic Series Expansion}\label{sec:intro_SSE}

Here we provide a brief outline of the thermal SSE for the Heisenberg AFM on the square lattice; see Ref.~\cite{sandvik1999stochastic} for a more comprehensive introduction. This will also provide us with an opportunity to establish notational conventions before introducing the post-measurement SSE. We will also discuss how the SSE can be used to rapidly sample sets of measurement outcomes according to the Born rule. 

The SSE is based on (i) an expansion of $e^{-\beta H}$ as a sum over operator strings, where each string is a product of operators contributing to $H$, and (ii) an importance sampling procedure for these operator strings. As our computational basis we choose eigenstates $\ket{\alpha}$ of all $S^z_j$ operators, writing the partition function as
\begin{equation}
   Z= \text{Tr}[e^{-\beta H}]=\sum_{\alpha}\sum_{n=0}^{\infty} \frac{(-\beta)^n}{n!} \braket{\alpha|H^{n}|\alpha}.
   \label{eqn:partition_function}
\end{equation}
Next, we decompose $H$ into a sum over terms that do not cause `branching' in the computational basis. 
This means that the action of each term in the expansion on a state $\ket{\alpha}$ generates another basis state $\ket{\alpha'}$, rather than a linear combination of such states. A standard choice is to decompose the two site Heisenberg operator $\bm{S}_j \cdot \bm{S}_k$ into its diagonal and off-diagonal parts \cite{sandvik_computational_2010}. We define
\begin{align}
    O^H_{1,(jk)}&=\frac{1}{2}\eye-2S^z_{j}S^z_{k}, \label{eqn:diag_with_constant}\\
    O^H_{2,(jk)}&=S_{j}^{+}S_{k}^{-}+S_{j}^{-}S_{k}^{+}, \notag
\end{align}
so that
\begin{equation}
  \bm{S}_j \cdot \bm{S}_k = \frac{1}{2} \left( O^H_{2,(jk)} - O^H_{1,(jk)} + \frac 1 2 \eye \right).
  \label{eq:heisenberg_decomp}
\end{equation}
Dropping the additive constant, the overall Hamiltonian can be expressed as
\begin{align}
  H=\sum_{d=1,2} (-1)^{d} \sum_{(jk)\in \text{NN}} \frac{J_{jk}}{2} \, O^H_{d,(jk)}.
\label{eqn:full_hamiltonian}
\end{align}
On inserting the decomposition of $H$ in Eq.~\eqref{eqn:full_hamiltonian} into Eq.~\ref{eqn:partition_function} we find a sum over strings $\mathcal{H}_{\bm{d}}$ of the operators $O^H_{d,(jk)}$. In practice we will work with a `fixed length' scheme. This amounts to fixing all operator strings to a length $L$ (guaranteed to always be larger than the expansion order $n$) by inserting additional identity operators \cite{sandvik_computational_2010}. Operator strings then have the form
\begin{align}
    \mathcal{H}_{\bm{d}} = \prod_{p=0}^{L-1} O^H_{d_p,(jk)_p},
    \label{eqn:opstring_H}
\end{align}
where the operator types can be $d_p=0$, $1$, or $2$, where we now use $d_p=0$ to refer to the identity operator $O^H_{0,(jk)}=\eye$. The index $\bm{d}$ for operator strings should be understood as an ordered list of operator types $d_p$ as well as the pairs of sites $(jk)_p$ on which these operators act. The number of type $d$ operator is $L_d$, such that $\sum_{d=0}^2 L_d = L$.

Having defined our operator strings we can assign statistical weights. The weight corresponding to basis state $\ket{\alpha}$ and operator string $\mathcal{H}_{\bm{d}}$ is $\mathcal{C}^H_{\bm{d}}\braket{\alpha|\mathcal{H}|\alpha}$ where
\begin{align}
    \mathcal{C}^H_{\bm{d}} = (\beta/2)^{L_1+L_2} (-1)^{L_2} \frac{L_0!}{L!} \prod_{p| d_p \neq 0} J_{(jk)_p},
    \label{eqn:stat_weight_H}
\end{align}
and the product of couplings $J_{(jk)_p}$ only runs over $d=1$ and $d=2$ type operators. The specific prefactors of $O^H_{d,(jk)}$ have been chosen so that the above matrix element is a boolean variable: $\braket{\alpha|\mathcal{H}|\alpha}=1$ for allowed configurations and $\braket{\alpha|\mathcal{H}|\alpha}=0$ otherwise. Note that on a bipartite lattice $\braket{\alpha|\mathcal{H}|\alpha} \neq 0$ only when $L_2$, the number of `hops', is even. Therefore, all contributing weights are non-negative. The combinatorial prefactor in $\mathcal{C}^H_{\bm{d}}$ is the ratio of (i) $1/(L_1+L_2)!$ from the Taylor expansion and (ii) ${L \choose L_1 + L_2}$, where (ii) compensates for the ${L \choose L_1 + L_2}$  ways of arranging non-identity operators in the length $L$ operator string. The full partition function in the fixed length scheme is
\begin{align}
    Z = \sum_{\alpha \bm{d}} \mathcal{C}^H_{\bm{d}} \braket{\alpha|\mathcal{H}_{\bm{d}}|\alpha}.
\end{align}
In order to sample basis states $\ket{\alpha}$ and operator strings $\mathcal{H}_{\bm{d}}$ according to the weights $\mathcal{C}^H_{\bm{d}} \braket{\alpha|\mathcal{H}|\alpha}$ we use two kinds of update, (i) `diagonal' moves, in which $d=0$ operators are replaced by $d=1$ operators, and vice versa, and (ii) `off-diagonal' moves in which $d=1$ and $d=2$ operators replace one another. For the latter we can use operator loop updates \cite{sandvik1999stochastic} (see Appendix~\ref{sec:loop_appendix}).

The different contributions to $Z$ can be visualized in terms of progagated states $\ket{\alpha(p)}$ with $p=0,\ldots,L-1$ defined modulo $L$. Each operator string $\mathcal{H}_{\bm{d}}$, here of the form in \cref{eqn:opstring_H}, can be factorized as $\mathcal{H}_{\bm{d}} = \mathcal{H}^{\geq p}_{\bm{d}}\mathcal{H}^{<p}_{\bm{d}}$, where $\mathcal{H}^{<p}_{\bm{d}}$ consists of the first $p$ operators in the string, and $\mathcal{H}^{\geq p}_{\bm{d}}$ consists of the final $L-p$. Here the propagated states have the form $\ket{\alpha(p)}=\mathcal{H}^{<p}_{\bm{d}}\ket{\alpha}$, with $\ket{\alpha}=\ket{\alpha(0)}=\ket{\alpha(L)}$ for $\braket{\alpha|\mathcal{H}_{\bm{d}}|\alpha}=1$. These expressions lead to an interpretation of the discrete index $p$ as a label for successive `slices' of the (continuous) imaginary time in the Euclidean spacetime path integral.

Before moving on to the post-measurement SSE, we discuss the problem of generating measurement outcomes according to the Born rule, i.e. `sampling' from a quantum state via measurement. Consider an experiment where $S^z_j$ is measured (projectively) on a set of sites $B$; the probability that our measurements return the outcomes $\ket{\alpha_B}$ is
\begin{align}
    \braket{\alpha_B|\text{Tr}_A[\rho] |\alpha_B} = Z^{-1}\sum_{\alpha_A \bm{d}} \mathcal{C}^H_{\bm{d}}\braket{\alpha|\mathcal{H}|\alpha} \label{eq:sampling}
\end{align}
where $\ket{\alpha}=\ket{\alpha_A,\alpha_B}$ is a basis state of the full system, with $A$ the complement of $B$. From this expression it is clear that the SSE automatically generates configurations of `measurement outcomes' $\ket{\alpha_B}$ according to the Born rule. This example is particularly simple because the measurement basis matches the computational basis. We now turn to the more general setting where these two bases are different, and we will also show how to calculate expectation values in the states generated by measurement. 

\section{Algorithm}\label{sec:algorithms}

In this section we show how to calculate correlation functions in `post-measurement' states generated by measurements on a large number of sites. We will focus on measurements which are \emph{not} diagonal in the computational basis, specifically measurements of the total spin $(\bm{S}_j +\bm{S}_k)^2$ on pairs of sites $(jk)$. The basic idea is to write the full post-measurement density matrix as a sum over operator strings, having contributions from $e^{-\beta H}$ as well as measurement operators, and to sample these operator strings. 

First, in \cref{sec:proj}, we formulate the post-measurement SSE for states created by large numbers of projective measurements. Following this, in \cref{sec:weak}, we develop an analogous toolkit for weak measurements.

\subsection{Projective measurements}\label{sec:proj}

Here we discuss projective measurement of the total spin $S^2_{(jk)} = \left( \bm{S}_j + \bm{S}_k \right)^2$ on pairs of sites $(jk)$. The set of pairs of sites that we measure is denoted $B$, and we restrict ourselves to the cases where $(jk)$ are nearest-neighbor (NN) or next-nearest-neighbor (NNN) sites on the square lattice. The two degenerate eigenspaces of $S_{(jk)}^2$ correspond to singlet and triplet states, having dimensions one and three, respectively. For a singlet outcome $s_{(jk)}=0$, our measurement disentangles the pair of sites $(jk)$ from all other degrees of freedom, while for a triplet outcome $s_{(jk)}=1$ we are left with an effective spin-$1$ degree of freedom on $(jk)$.

The projection operators $P^{s_{(jk)}}_{(jk)}$ corresponding to these outcomes are given in Eqs.~\eqref{eqn:proj_defn_singlet} and \eqref{eqn:proj_defn_triplet}. For a set of outcomes $\bm{s}=\{s_{(jk)}| (jk) \in B\}$ the product of all of the projectors $P^{\bm{s}}=\prod_{(jk)\in B} P^{s_{(jk)}}_{(jk)}$ generates the post-measurement density matrix $\rho^{\bm{s}} \propto P^{\bm{s}}\rho P^{\bm{s}}$. The post-measurement SSE will allow us to evaluate post-measurement expectation values of observables $A$,
\begin{align}
    \braket{A}^{\bm{s}} = \frac{\text{Tr}[P^{\bm{s}} e^{-\beta H} P^{\bm{s}}A]}{\text{Tr}[P^{\bm{s}} e^{-\beta H} P^{\bm{s}}]}. \label{eq:postmeasurementexpectation}
\end{align}
Like the thermal SSE described above, our algorithm will be based on an expansion of the partition function
\begin{align}
    Z^{\bm{s}} = \text{Tr}[P^{\bm{s}} e^{-\beta H} P^{\bm{s}}]\label{eq:Zs}
\end{align}
as a sum over operators that do not cause branching of $S^z_j$ basis states. 

Before proceeding to the algorithm, we note that the evaluation of a post-measurement expectation value is an elementary step when sampling a full set $\bm{s}$ of measurement outcomes according to the Born rule. An algorithm to evaluate post-measurement expectation values is, by construction, also an algorithm to sample large sets of measurement outcomes. To see this consider first measuring a pair of sites $(jk)$; the probability to measure $s_{(jk)}$ is simply $\text{Tr}[\rho P^{s_{(jk)}}_{(jk)}]$. If we find outcome $s_{(jk)}$, we generate a post-measurement state $\rho^{s_{(jk)}} \propto P^{s_{(jk)}}_{(jk)} \rho P^{s_{(jk)}}_{(jk)}$. Next, we measure another pair of sites $(j'k')$; the probability to find outcome $s_{(j'k')}$ is the expectation value of $P^{s_{(j'k')}}_{(j'k')}$ with respect to $\rho^{s_{(jk)}}$. Iterating this procedure, we can sample a full set of measurement outcomes according to the Born rule. For now we will nevertheless focus on post-measurement expectation values in specific states. 

To construct the post-measurement SSE we write the decomposition of individual projection operators described in Eqs.~\ref{eqn:proj_defn_singlet} and \ref{eqn:proj_defn_triplet} as
\begin{align}
    P^{0}_{(jk)} &= \frac{1}{2}O^0_{1,(jk)}-\frac{1}{2}O^0_{2,(jk)}, \label{eqn:proj_decomposition}\\
    P^{1}_{(jk)} &= \frac{1}{2}O^1_{0,(jk)} +  \frac{1}{2}O^1_{1,(jk)} +  \frac{1}{2}O^1_{2,(jk)},\notag
\end{align}
where the superscript indicates the outcome for the total spin $s_{(jk)}$ on the measured bond ($0$ for singlet and $1$ for triplet). In the singlet case the operators are
\begin{align}
    O^0_{1,(jk)}=\frac{1}{2}-2S^z_j S^z_k, \quad
    O^0_{2,(jk)}&=S^+_j S^-_k + S^-_jS^+_k, 
\label{eq:singletdecomposition}
\end{align}
while in the triplet case $O^1_{0,(jk)} = \eye$ and
\begin{align}
    O^1_{1,(jk)}=\frac{1}{2}+2S^z_j S^z_k, \quad
    O^1_{2,(jk)}=S^+_j S^-_k + S^-_jS^+_k.
\label{eq:tripletdecomposition}
\end{align}
Although in the problem we study here the operators appearing in the projection operators have a similar form as those appearing in $H$, this is not the case in general.

On inserting these expansions into the products of projection operators defining the post-measurement density matrix, we find 
\begin{align}
    P^{\bm{s}} = \sum_{\bm{d}} \mathcal{C}^{\bm{s}}_{\bm{d}} \mathcal{P}^{\bm{s}}_{\bm{d}},
\end{align}
where $\mathcal{C}^{\bm{s}}_{\bm{d}}$  is a scalar and 
\begin{align}
    \mathcal{P}^{\bm{s}}_{\bm{d}} = \prod_{(jk) \in B} O^{s_{(jk)}}_{d_{(jk)},(jk)}
\end{align}
is an operator string. In each operator string $\mathcal{P}^{\bm{s}}_{\bm{d}}$ corresponding to the set of outcomes $\bm{s}$, there is exactly one operator $O^{s_{(jk)}}_{d_{(jk)},(jk)}$ coming from each pair of spins $(jk) \in B$. As above, the notation $\bm{d}$ is a list of operator types. The explicit values of the scalars $\mathcal{C}^{\bm{s}}_{\bm{d}}$  can be read off of our expressions above: $\mathcal{C}^{\bm{s}}_{\bm{d}} = 2^{-|B|}\prod_{(jk) \in B} (-1)^{\delta_{0,s_{(jk)}}\delta_{2,d_{(jk)}}}$, where $|B|$ is the number of measurements.  

\begin{figure}[h!]
    \centering
    \includegraphics[width=0.44\textwidth]{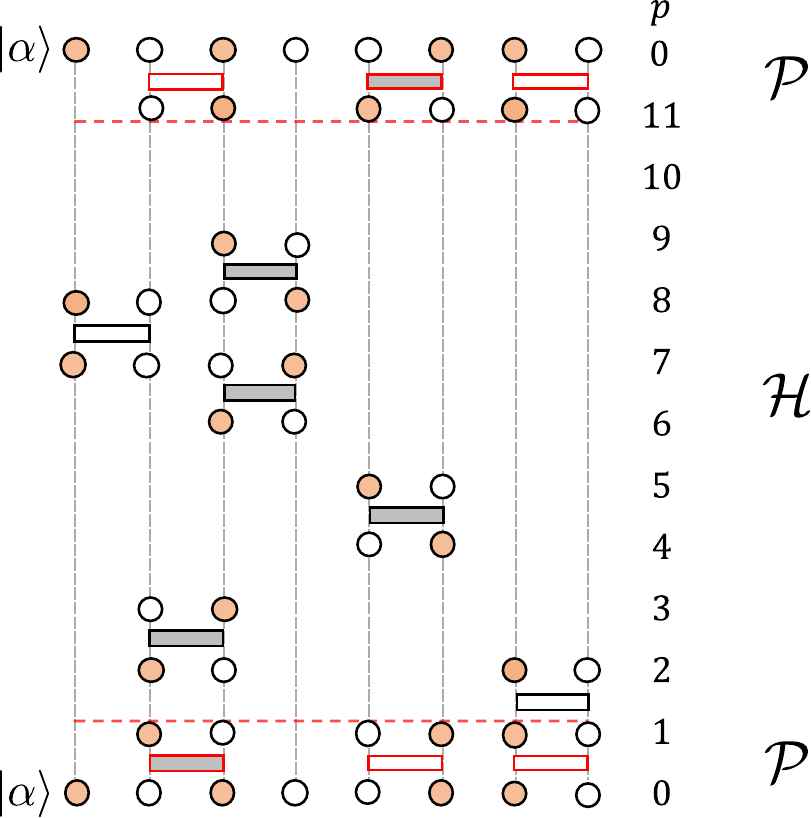}
    \caption{Example post-measurement SSE configuration for a spin chain, with all propagated states shown. Dashed lines indicate that the propagated state on a site is unchanged with respect to the previous step. Open (filled) circles denote $\ket{\downarrow}$ ($\ket{\uparrow}$) states, while open (solid) bars represent diagonal $O_{1,(jk)}$ (off-diagonal $O_{2,(jk)}$) operators, and the absence of a bar represents an identity operator. The configuration displayed consists of $L_1 = 6$ diagonal operators located after $p = 0,1,7, 11$ and $L_2 = 6$ off-diagonal operators after $p = 0, 2, 4, 6, 8, 11$. The post-measurement correlations of diagonal observables are evaluated, with respect to $\ket{\alpha(p=0)}=\ket{\alpha}$, in the $p=0$ imaginary time slice.}
  \label{fig:SSE_example_config}
\end{figure}

The resulting expression for the partition function in Eq.~\eqref{eq:Zs} is
\begin{align}
    Z^{\bm{s}} = \sum_{\alpha\bm{a} \bm{b} \bm{c}} \mathcal{C}^H_{\bm{a}} \mathcal{C}^{\bm{s}}_{\bm{b}}\mathcal{C}^{\bm{s}}_{\bm{c}}  \braket{\alpha|\mathcal{P}^{\bm{s}}_{\bm{b}}\mathcal{H}_{\bm{a}}\mathcal{P}^{\bm{s}}_{\bm{c}}|\alpha}. \label{eqn:Zsstrings}
\end{align}
The post-measurement SSE is based on Eq.~\eqref{eqn:Zsstrings}. Note that the operator string $\mathcal{P}_{\bm{b}}^{\bm{s}}\mathcal{H}_{\bm{a}}\mathcal{P}_{\bm{c}}^{\bm{s}}$, which has length $L+2|B|$, does not cause branching of basis states. From our definitions of $O^{s_{(jk)}}_{d,(jk)}$, the matrix elements $\braket{\alpha|\mathcal{P}_{\bm{b}}^{\bm{s}}\mathcal{H}_{\bm{a}}\mathcal{P}_{\bm{c}}^{\bm{s}}|\alpha}=0,1$ remain boolean variables, and the coefficient $\mathcal{C}^H_{\bm{a}}$ has the same form as in the standard SSE discussed in the previous section. Propagated states $\ket{\alpha(p)}$ can here be defined, in analogy with those appearing in the thermal SSE, as the basis states generated by acting with successive operators in $\mathcal{P}^{\bm{s}}_{\bm{b}}\mathcal{H}_{\bm{a}}\mathcal{P}^{\bm{s}}_{\bm{c}}$ on $\ket{\alpha}$, and we illustrate an example configuration in Fig.~\ref{fig:SSE_example_config}.

When evaluating post-measurement expectation values, observables must be inserted between the two measurement strings. For example, for an observable $A$ that is diagonal in the computational basis, e.g. $A = S^z_j S^z_k$, we have 
\begin{align}
    \braket{A}^{\bm{s}} = \frac{1}{Z^{\bm{s}}}\sum_{\alpha \bm{a}\bm{b}\bm{c}}\mathcal{C}^H_{\bm{a}} \mathcal{C}^{\bm{s}}_{\bm{b}}\mathcal{C}^{\bm{s}}_{\bm{c}}  \braket{\alpha|\mathcal{P}^{\bm{s}}_{\bm{b}}\mathcal{H}_{\bm{a}}\mathcal{P}^{\bm{s}}_{\bm{c}}|\alpha}\braket{\alpha|A|\alpha}. 
\end{align}
We will focus on expectation values of operators that are \SU{2} symmetric, in particular $\bm{S}_j \cdot \bm{S}_k$. This operator is not diagonal in the computational basis but, since $\rho^{\bm{s}}$ is \SU{2} symmetric, we have $\braket{\bm{S}_j\cdot \bm{S}_k}^{\bm{s}} = 3 \braket{S^z_j S^z_k}^{\bm{s}}$. We can therefore restrict ourselves to diagonal observable $A$. 

As noted above, expectation values must be evaluated at imaginary time `$\tau=0$', i.e. between the two measurement strings as indicated in Fig.~\ref{fig:SSE_example_config}. This situation should be contrasted with the thermal SSE, where imaginary-time translation symmetry of operator strings means that expectation values can be evaluated as averages over `time slices' $\tau$. This difference arises from the fact that measurements explicitly break the time-translation symmetry of the imaginary-time path integral.

\subsection{Weak measurements}\label{sec:weak}

Here we discuss how our scheme can be generalized to study the effects of Kraus operators that are not simple projectors. Our focus will be on weak measurements, which approach projective measurements only in the limit where the `measurement strength' $\mu$ is increased to infinity.

The weak measurement of a Heisenberg interaction on the pair of sites $(jk)$ corresponds to the action of one of the Kraus operators
\begin{align}
    W^s_{(jk)} = w_s e^{(s-1/2)\mu \bm{S}_j \cdot \bm{S}_k}
\end{align}
where $s=0,1$. The real constants $w_s \geq 0$ are uniquely determined by the normalization condition ${\sum_s [W^s_{(jk)}]^2=\eye}$. 

We now seek an expression for the partition function which generates correlations in the post-measurement density matrix $\rho^{\bm{s}}_\mu$; away from the projective ($\mu=\infty$) and unmeasured ($\mu=0$) limits we keep the $\mu$-dependence of the density matrix explicit. Weak measurements on the set of pairs of sites $B$, having outcomes $\bm{s}$ and uniform $\mu$, create
\begin{align}
    \rho^{\bm{s}}_{\mu} \propto W^{\bm{s}}_{(jk)} e^{-\beta H} W^{\bm{s}}_{(jk)},
\end{align}
where we have defined $W^{\bm{s}}=\prod_{(jk) \in B} W^{s_{(jk)}}_{(jk)}$. Here the $\mu$-dependence of the measurement operators is implicit, and we have omitted the normalization. The corresponding partition function when starting from a thermal state $\rho \propto e^{-\beta H}$ is then
\begin{align}
    Z^{\bm{s}}_{\mu} = \text{Tr}\big[ \big(W^{\bm{s}}\big)^2 e^{-\beta H}].
\end{align}
As in the case of projective measurements, the next step is to expand each of $H$ and $W^{\bm{s}}$ as sums over operator strings which do not cause branching of computational basis states. 

The weak measurement algorithm can be implemented along similar lines to the projective measurement algorithm. To show this we write
\begin{align}
    W^{s}_{(jk)} \propto [e^{\mu/2}-1]^{-1} \eye + P^{s}_{(jk)},
\end{align}
and then expand $P^s_{(jk)}$ as in the previous section. The results are
\begin{align}
    W^{0}_{(jk)} &\propto [e^{\mu/2}-1]^{-1}O^{0}_{0,(jk)} + \frac{1}{2}O^{0}_{1,(jk)} - \frac{1}{2}O^{0}_{2,(jk)},\label{eq:Wsdecomposition}\\
    W^{1}_{(jk)} &\propto [2\tanh(\mu/4)]^{-1} O^{1}_{0,(jk)} + \frac{1}{2}O^{1}_{1,(jk)} + \frac{1}{2}O^{1}_{2,(jk)},\notag
\end{align}
where now the $d=0$ operator $O^0_{0,(jk)}=\eye$ appears for $s=0$. For $\mu=0$ the `weak measurements' reduce to identity operators. Inserting decompositions of $W^s_{(jk)}$ of this kind into each factor of $W^{\bm{s}}_{(jk)}$ appearing in the partition function we are left with a sum over operator strings $\mathcal{P}^{\bm{s}}_{\bm{d}}$. These operator strings have the same form as for projective measurements, but for finite $\mu$ their weights $\mathcal{C}^{\bm{s}}_{\bm{d}, \mu}$ can now be nonzero when $d=0$ on a pair of sites with outcome $s=0$.

From this step on, the analysis is essentially identical to projective measurements. Inserting the decompositions of operator strings we find that the partition function $Z^{\bm{s}}_{\mu}$ differs from $Z^{\bm{s}}$ ($=Z^{\bm{s}}_{\infty}$) in Eq.~\eqref{eqn:Zsstrings} only by the replacement of coefficients $\mathcal{C}^{\bm{s}}_{\bm{d}}$  ($=\mathcal{C}^{\bm{s}}_{\bm{d},\infty}$) with $\mathcal{C}^{\bm{s}}_{\bm{d}\mu}$. The values of these coefficients can be determined from Eq.~\eqref{eq:Wsdecomposition}; for example, in the cases where $\bm{s}=\bm{0}$ and $\bm{s}=\bm{1}$ on all measured pairs of sites, these coefficients are respectively equal to 
\begin{align}
    \mathcal{C}_{\bm{d},\mu}^{\bm{0}} &= 2^{L_0}[e^{\mu/2}-1]^{-L_0} (-1)^{L_2},\\
    \mathcal{C}^{\bm{1}}_{\bm{d},\mu} &= [\tanh(\mu/4)]^{-L_0},\notag
\end{align}
up to unimportant overall prefactors. Here $L_d$ denotes the number of type $d$ operators in the string $\bm{d}$ referenced on the left-hand side of the equation.

For consistency, we have here described the weak measurement scheme along similar lines to the projective measurement scheme. Note, however, that we can also write $W^{\bm{s}} \propto e^{-\mu M^{\bm{s}}/2}$, where $M^{\bm{s}}$ is a sum of local terms. In this notation the partition function involves a product of $e^{-\mu M^{\bm{s}}}$ and $e^{-\beta H}$. Just as $e^{-\beta H}$ is expanded in powers of $\beta$, the operator $e^{-\mu M^{\bm{s}}}$ can be expanded in powers of $\mu$. This leads to an algorithm where we view $M^{\bm{s}}$ as the generator of `imaginary time' evolution for an interval $\mu$. We detail this algorithm in \Cref{sec:weak_meas_appendix}, and use this in practice for our finite $\mu$ numerical calculations. Note that this algorithm allows for a general description of correlations in systems described by a combination of different imaginary time evolutions. For example, it also allows for the investigation of strange correlators \cite{you2014wave}, which were previous calculated using QMC in Ref~\cite{wu2015quantum}.

\begin{figure}[ht]
\includegraphics[width=0.44\textwidth]{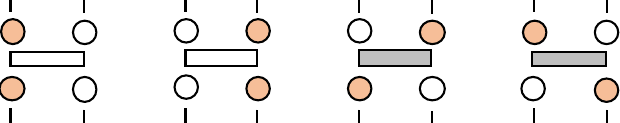}\\
\vspace{0.2in}
\includegraphics[width=0.44\textwidth]{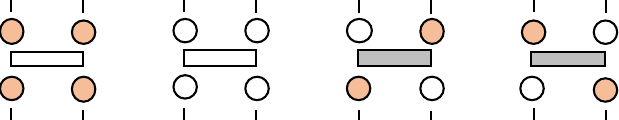}
\caption{Allowed vertices in the measurement string for (upper row) $s=0$ outcomes and (lower row) $s=1$ outcomes. Open and filled circles represent $\ket{\downarrow}$ and $\ket{\uparrow}$ states, respectively, while open and filled boxes represent (non-identity) diagonal $O^s_{1,(jk)}$ and off-diagonal $O^s_{2,(jk)}$ operators. Diagonal operators act on anti-parallel spins for $s=0$, whereas they act on parallel spins for $s=1$.}
\label{fig:allowed_vertices}
\end{figure}
\subsection{Numerical implementation}\label{sec:implementation}

Here we discuss how to use the above expansions of $Z^{\bm{s}}$ to evaluate post-measurement expectation values in practice. We then describe the sets of measurement outcomes for which $Z^{\bm{s}}$ is sign free, and benchmark our algorithm against exact diagonalization (ED) in small systems. For simplicity we restrict to projective measurements. 

For a fixed set of outcomes $\bm{s}$, the sampling algorithm starts from a basis state $\ket{\alpha}$ as well as the three parts of the operator string $\mathcal{H}_{\bm{a}}$, $\mathcal{P}^{\bm{s}}_{\bm{b}}$, and $\mathcal{P}^{\bm{s}}_{\bm{c}}$, such that the overall configuration is allowed $\braket{\alpha|\mathcal{P}^{\bm{s}}_{\bm{b}}\mathcal{H}_{\bm{a}}\mathcal{P}^{\bm{s}}_{\bm{c}}|\alpha}=1$. The basic idea is to propose changes in the operator string and the basis state which take us between different allowed configurations, accepting changes according to their statistical weights (as in any standard Metropolis-Hastings algorithm). 

To describe our implementation in more detail it is necessary to specify the measurement outcomes $\bm{s}$. Two classes of outcomes $\bm{s}$ for which the statistical weights $\mathcal{C}^H_{\bm{a}} \mathcal{C}^{\bm{s}}_{\bm{b}}\mathcal{C}^{\bm{s}}_{\bm{c}}$ are positive are as follows:

\paragraph{NN singlets:}
For $B$ consisting only of non-overlapping NN sites, and for outcomes $\bm{s}=0$, the sign of the weight is the parity of the number of type $2$ operators appearing in $\mathcal{P}^{\bm{s}}_{\bm{b}}\mathcal{H}_{\bm{a}}\mathcal{P}^{\bm{s}}_{\bm{c}}$. The condition $\braket{\alpha|\mathcal{P}^{\bm{s}}_{\bm{b}}\mathcal{H}_{\bm{a}}\mathcal{P}^{\bm{s}}_{\bm{c}}|\alpha}=1$ implies that this number is even. 
\paragraph{NNN triplets:} 
For $B$ a set of non-overlapping NNN sites, and for outcomes $\bm{s}=1$, the sign of the weight is the parity of the number of type $2$ operators appearing in $\mathcal{H}_{\bm{a}}$ alone. Spin flips generated by $O^H_{(jk),2}$ in $\mathcal{H}_{\bm{a}}$ are along NN $(jk)$ and are associated with negative signs, while spin flips generated by $O^1_{(jk)}$ operators in $\mathcal{P}^{\bm{1}}_{\bm{b}}$ and $\mathcal{P}^{\bm{1}}_{\bm{c}}$ are along NNN $(jk)$ and come with positive signs. The condition $\braket{\alpha|\mathcal{P}^{\bm{s}}_{\bm{b}}\mathcal{H}_{\bm{a}}\mathcal{P}^{\bm{s}}_{\bm{c}}|\alpha}=1$ then implies that the number of type $2$ operators in $\mathcal{H}_{\bm{a}}$ is even. 

This discussion also makes clear that any combination of singlets on NNs and triplets on NNNs is also sign-problem-free, and so can be studied using the post-measurement SSE formulated in the basis $\ket{\alpha}$. 

We now describe our algorithms for sampling operator strings. For `singlet measurements' ($\bm{s}=\bm{0}$ with $B$ a set of NNs) we initialize the $\mathcal{P}^{\bm{0}}_{\bm{b}}$ and $\mathcal{P}^{\bm{0}}_{\bm{c}}$ operator strings with $\bm{b}=\bm{c}=\bm{1}$, i.e. type $1$ operators on all NNs. Recall that, in this setup, we never encounter type $0$ operators in the singlet projectors. Therefore, we don't need to perform `diagonal' moves which exchange type $0$ and type $1$ operators in the measurement strings. Instead, in the measurement strings, we only need to focus on off-diagonal updates (which exchange type $1$ and type $2$ operators). Since here the diagonal (type $1$) and off-diagonal (type $2$) operators have weights of the same magnitude, it becomes possible to use highly efficient loop updates \cite{syljuassen2002quantum} of the entire $\mathcal{P}^{\bm{0}}_{\bm{b}}\mathcal{H}_{\bm{a}}\mathcal{P}^{\bm{0}}_{\bm{c}}$ operator string. Such updates satisfy detailed balance and are designed to always give allowed operator strings.

\begin{figure}[ht]
    \includegraphics[width=0.45\textwidth]{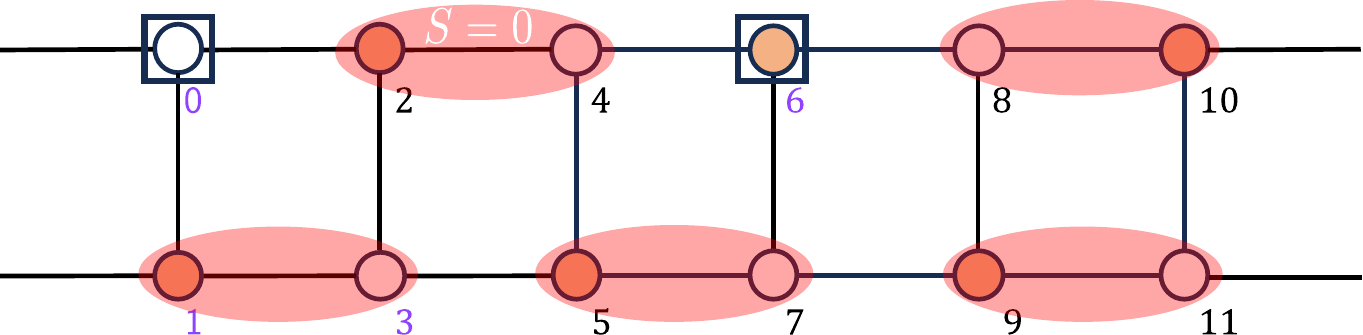}\\
    \vspace{0.2in}
    \includegraphics[width=0.45\textwidth]{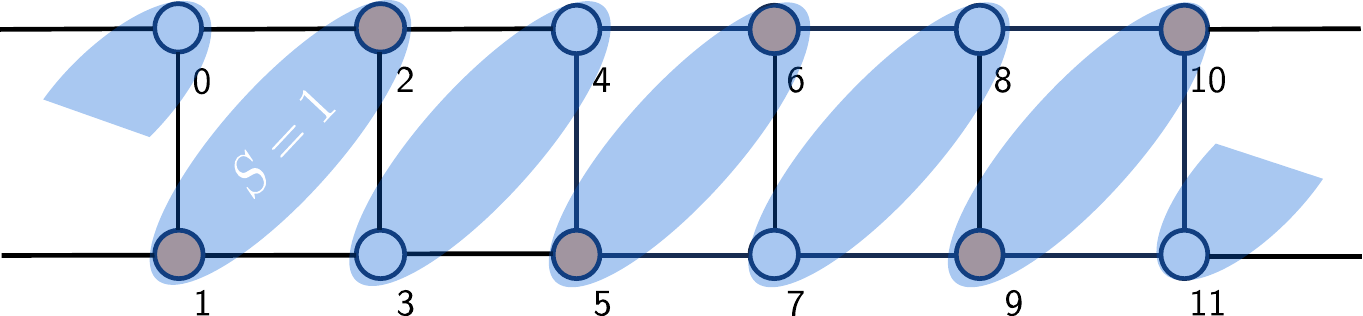}
\caption{Post-measurement states of the $2 \times 6$ spin ladder used for benchmarking in Fig.~\ref{fig:proj_corr_heisenberg_ladder}. Solid black lines indicate NN interactions in the Hamiltonian. (Upper) $\bm{s}=\bm{0}$ outcomes on a nonoverlapping set of NNs, shown as red ellipses. The two spins indicated using black boxes are left unmeasured. (Lower) $\bm{s}=\bm{1}$ outcomes on the nonoverlapping set of NNNs, shown as blue ellipses.}
    \label{fig:benchmarking_models}
\end{figure}

\begin{figure}
    \centering
    \includegraphics{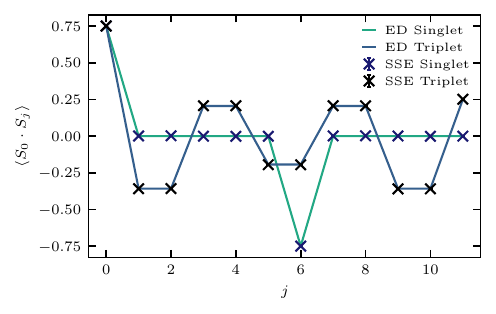}
    \caption{Benchmarking post-measurement SSE against ED for a $2\times 6$ 
lattice. We study post-measurement states arising from both singlet and triplet outcomes as depicted in the upper and lower panels of Fig.~\ref{fig:benchmarking_models}, respectively. The inverse temperature $\beta=18$, and $J_1=J_2=1$.}
    \label{fig:proj_corr_heisenberg_ladder}
\end{figure}

For `triplet measurements' ($\bm{s}=\bm{1}$ with $B$ a set of NNNs) we start with $\mathcal{P}_{\bm{b}}$ and $\mathcal{P}_{\bm{c}}$ consisting of type $0$ operators (i.e. identity operators) and then perform both diagonal ($d=0 \leftrightarrow d=1$) and off-diagonal  ($d=1 \leftrightarrow d=2$) updates in these operator strings. Here, when implementing the deterministic loop update, it is important to account for the fact that the diagonal operators $O^H_{1,(jk)}$ and $O^{\bm{1}}_{1,(jk)}$, appearing in $\mathcal{H}_{\bm{a}}$ and the measurement strings, respectively, annihilate different spin configurations: $O^H_{1,(jk)}\ket{\uparrow_j \uparrow_k}=O^H_{1,(jk)}\ket{\downarrow_j \downarrow_k}=0$ while $O^{\bm{1}}_{1,(jk)}\ket{\uparrow_j \downarrow_k}=O^{\bm{1}}_{1,(jk)}\ket{\downarrow_j \uparrow_k}=0$. The different spin and operator configurations allowed in the SSE are illustrated in Fig.~\ref{fig:allowed_vertices}.

As a numerical demonstration, we now study the effects of projective singlet and triplet measurements in small systems where we can benchmark our results against exact diagonalization (ED). The two measurement configurations are illustrated in Fig.~\ref{fig:benchmarking_models}. To test our implementation of $\bm{s}=\bm{0}$ measurements we force $s_{(jk)}=0$ on a non-overlapping set of NNs which cover all but two separated spins. Since at large $\beta$ the initial state $\rho \sim e^{-\beta H}$ approaches the ground state, which is a total singlet, these measurements force the \textit{unmeasured} spins into a non-local singlet. Therefore $\braket{S^z_j S^z_k}^{\bm{0}} = \braket{\bm{S}_j \cdot \bm{S}_k}^{\bm{0}}/3 \to -1/4$ as $\beta \to \infty$. To test our implementation of $\bm{s}=\bm{1}$ measurements we force $s_{(jk)}$ on a non-overlapping set of NNNs which cover every spin; this generates a set of $N/2$ spin-$1$ degrees of freedom whose correlations are inherited from the initial state. The results of our numerical calculations are shown in Fig.~\ref{fig:proj_corr_heisenberg_ladder}, and in each case we find quantitative agreement between SSE and ED. In the next section we will study both weak and projective measurements; our results there demonstrate the accuracy of our weak measurement algorithm. 
\section{Symmetry of measured states}\label{sec:singlet}

The ground states of Heisenberg AFMs on bipartite lattices are total spin singlets. Our measurements preserve the \SU{2} symmetry of this state, so it is natural to ask when this symmetry has interesting consequences for post-measurement correlations. For example, if we find singlets $\bm{s}=\bm{0}$ covering all but two unmeasured spins, those spins must form a singlet; see Fig.~\ref{fig:proj_corr_heisenberg_ladder}. First, in Sec.~\ref{sec:ordered}, we discuss the role of the \SU{2} symmetry in ordered phase. We then discuss the dimerized phase in Sec.~\ref{sec:disordered}.
\subsection{Ordered phase}\label{sec:ordered}
Although the ground state of our model is a total spin singlet, at any nonzero temperature the thermal density matrix has only a weak \SU{2} symmetry. Moreover, in the ordered phase (which includes the isotropic Heisenberg model), there is a `tower of states' above the singlet ground state. While the low-energy spin waves have a linear dispersion, and hence in a finite system have energies $\sim N^{-1}_x$, the gap $\Delta$ to the tower of states is expected to scale as $\Delta \sim N^{-1} \sim N_x^{-2}$ \cite{auerbach2012interacting}. Since our algorithm works directly with thermal states, these excitations will have important consequences for the effects of measurements. 

\begin{figure}[!ht]
\includegraphics{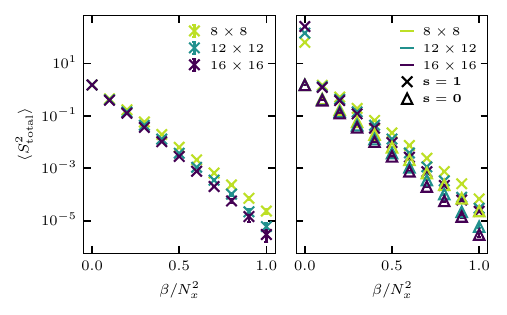}
\caption{Total spin $\langle S^2_{\text{total}}\rangle$ as a function of $\beta/N_x^2$ for various $N_x = N_y$ (legend) in (left) the unmeasured thermal state $\rho \sim e^{-\beta H}$ and (right) the post-measurement states $\rho^{\bm{0}}$ (triangles) and $\rho^{\bm{1}}$ (crosses). Here $\rho^{\bm{0}}$ is created by measurements of singlets $\bm{s}=\bm{0}$ on a set of NNs excluding two sites separated by $N_x/2$ along the horizontal, and $\rho^{\bm{1}}$ is created by measurements of triplets $\bm{s}=\bm{1}$ on a full set of non-overlapping NNNs [see e.g. Fig.~\ref{fig:aklt}]. Here $J_1=J_2=1$.}.
\label{fig:tot_spin_with_diff_latt_sizes}
\end{figure}

We first characterize the symmetry of unmeasured thermal states in the isotropic Heisenberg model by evaluating the expectation value of the total spin $\braket{S^2_{\text{total}}}$. Invoking the (weak) \SU{2} symmetry of the thermal state, we can calculate this quantity as $\braket{S^2_{\text{total}}}=3\sum_{jk}\braket{S^z_jS^z_k}$. The right-hand side of this expression is straightforward to evaluate using the thermal SSE since it is a sum over correlation functions of diagonal operators $S^z_j$. At large $\beta$ these thermal expectation values have significant contributions only from the Anderson tower. The first excited state in the tower is expected to have a total spin of unity \cite{auerbach2012interacting}, and hence is threefold degenerate. Therefore, at large $\beta$,
\begin{align}
    \braket{S^2_{\text{total}}} = 6 e^{-\beta \Delta} +\ldots,
\end{align}
and we verify this behavior with $\Delta \sim N_x^{-2}$ in the left panel of \cref{fig:tot_spin_with_diff_latt_sizes}.

When we measure the system, the relative contributions of different energy eigenstates to the density matrix are modified, and it is important to understand which eigenstates dominate post-measurement correlations. To address this, we will evaluate $\braket{ S^2_{\text{total}}}^{\bm{s}}$ in post-measurement states $\rho^{\bm{s}}$. Recall that all eigenstates $\ket{n}$ of the Hamiltonian $H$ are also eigenstates of the total spin, and that any eigenstate with spin $S_n$, i.e. with $S^2_{\text{total}}\ket{n}=S_n(S_n+1)\ket{n}$, belongs to a $(2S_n+1)$-fold degenerate multiplet. Moreover, since the measurement operators $P^{\bm{s}}$ are \SU{2} symmetric, they do not mix sectors having different $S^z_{\text{total}}$ or $S^2_{\text{total}}$ quantum numbers. Denoting by $p^{\bm{s}}_n = \braket{n|P^{\bm{s}}|n}$ the probability to observe outcomes $\bm{s}$ given eigenstate $\ket{n}$, the post-measurement expectation value of the total spin $\braket{S^2_{\text{total}}}^{\bm{s}} = \text{Tr}[P^{\bm{s}} e^{-\beta H} P^{\bm{s}} S^2_{\text{total}}]/\text{Tr}[P^{\bm{s}}e^{-\beta H}]$ is given by
\begin{align}
    \braket{S^2_{\text{total}}}^{\bm{s}} &= \frac{\sum_n e^{-\beta E_n} p^{\bm{s}}_n S_n(S_n+1)}{\sum_n e^{-\beta E_n} p^{\bm{s}}_n},
\end{align}
where $H\ket{n}=E_n\ket{n}$. Here we have used $\braket{n|P^{\bm{s}}S^2_{\text{total}}P^{\bm{s}}|n}=S_n(S_n+1) p^{\bm{s}}_n$, which follows from $[P^{\bm{s}},S^2_{\text{total}}]=0$. At large inverse temperature $\beta$ we then find
\begin{align}
     \braket{S^2_{\text{total}}}^{\bm{s}} = 6(p^{\bm{s}}_1/p^{\bm{s}}_0) e^{-\beta \Delta} + \ldots, \label{eq:totalspinmeasured}
\end{align}
where we have again used the fact that the first excited states have total spin of unity. The ellipsis denotes contributions from higher-energy states. 

The appearance of a ratio of Born probabilities $p^{\bm{s}}_1/p^{\bm{s}}_0$ in Eq.~\eqref{eq:totalspinmeasured} is concerning: each of the probabilities $p^{\bm{s}}_{0}$ and $p^{\bm{s}}_{1}$ are products of an extensive number of (conditional) probabilities for the outcomes of local measurements, so it would be natural to expect $p^{\bm{s}}_n$ to have a very broad distribution, exhibiting fluctuations over $n$ which diverge with increasing numbers of measurements. If this was the case, we could easily find ourselves in a scenario where $p^{\bm{s}}_1/p^{\bm{s}}_0$ is very large, and hence where $\rho^{\bm{s}}$ is dominated by excited eigenstates. 

This scenario does not appear to arise in practice, at least for the measurement outcomes that we consider. In many systems with local interactions, in particular those which can be described in terms of quasiparticle at low energies, the ground and lowest excited states generally resemble one another at the level of local expectation values. This `local similarity' of low-lying states would suggest that the Born probabilities $p^{\bm{s}}_n$ for fixed $\bm{s}$ exhibit only moderate variations with $E_n$, at least for $E_n$ close to the ground state energy, and hence that ratios of probabilities such as $p^{\bm{s}}_1/p^{\bm{s}}_0$ do not diverge with system size. The results in Fig.~\ref{fig:tot_spin_with_diff_latt_sizes} are consistent with this expectation: we calculate $\braket{S^2_{\text{total}}}^{\bm{s}}$ for (i) $\bm{s}=\bm{0}$ on a set of NNs excluding two spins separated by $N_x/2$ along the horizontal, (ii) for $\bm{s}=\bm{1}$ on a full set of NNNs. The scaling of the total spin with inverse temperature and system size allows us to identify the regime where the post-measurement state is dominated by the ground state, and it is clear that $p^{\bm{s}}_1/p^{\bm{s}}_0$ is of order unity. The behavior identified in Figs.~\ref{fig:tot_spin_with_diff_latt_sizes} indicates that we need only use $\beta \gg \Delta^{-1} \sim N$ in order to probe the \SU{2}-symmetric ground state. 

Moving away from the ground state, in the following we will set $\beta = 3N_x$. Since the spin wave speed in the isotropic model is $1.67(3)$ (where $J_1$, $J_2$, $\hbar$, and the lattice constant are all unity) \cite{singh1989microscopic} and the longest wavelength is $2\pi/N_x$ in a system with periodic boundary conditions, setting $\beta = 3N_x$ ensures that our thermal density matrix $\rho \sim e^{-\beta H}$ is well-approximated by the contributions from the tower of states. Therefore, for large $N_x$, it is reasonable to expect $\rho$ to behave like a statistical mixture of classical Ne\'{e}l-ordered ground states.

To test this idea, we return to the setting of \cref{fig:proj_corr_heisenberg_ladder}, i.e. to the creation of a non-local singlet by forcing $\bm{s}=\bm{0}$ on NN pairs of spins (red ellipses in \cref{fig:proj_corr_heisenberg_ladder}), leaving two well-separated spins unmeasured. For $\beta \Delta \gg 1$ the post-measurement correlations are $\braket{\bm{S}_j \cdot\bm{S}_k}^{\bm{0}}=-3/4$, while at infinite temperature $\beta=0$ the correlations vanish. The picture of $\rho \sim e^{-\beta H}$ (with $\beta \sim  N_x$) as a statistical mixture of Ne\'{e}l-ordered states related by rotation suggests that the magnitude of $\braket{\bm{S}_j \cdot\bm{S}_k}^{\bm{0}}$ should decrease as $N_x$ is increased. 

\begin{figure}[!ht]
\includegraphics{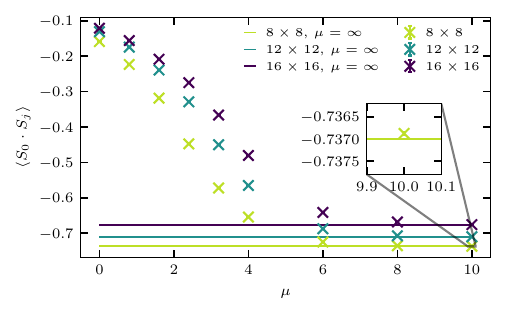}
\caption{Correlations between unmeasured spins following weak and projective measurements with $\bm{s}=\bm{0}$ on a set of NN pairs of sites excluding $0$ and $j$. As the strength of weak measurement $\mu$ is increased, the post-measurement correlations (crosses) converge to the result for projective measurements (solid horizontal lines); see inset. Here $N_x=N_y$ (legend), $\beta=3\times N_x$, and $J_1=J_2=1$.}
\label{fig:unmeas_spin_cor_vs_mu}
\end{figure}

This behavior is clear in Fig.~\ref{fig:unmeas_spin_cor_vs_mu}, where we also test our weak measurement algorithm. We calculate $\braket{\bm{S}_j \cdot\bm{S}_k}^{\bm{0}}$ for various measurement strengths $\mu$, where the expectation value is with respect to $\rho^{\bm{0}}_{\mu}$. The spatial arrangement of singlet measurements is as depicted in the upper diagram in Fig.~\ref{fig:benchmarking_models}, where the two unmeasured spins are maximally separated along the horizontal direction and with the pattern of singlet measurements repeating for all additional rows (i.e. alternating between horizontal singlets on even bonds and horizontal singlets on odd bonds). Our results show that, on increasing the measurement strength $\mu$, the correlation function $\braket{\bm{S}_j\cdot\bm{S}_k}^{\bm{0}}$ between the unmeasured spins decreases from its ground state value to near $-3/4$, the value for a long-range singlet. However, as the system size is increased with $\beta = 3N_x$, the post-measurement correlations are indeed weakened.

The situation identified above for ordered phases is to be contrasted with that in symmetric (i.e. disordered) phases. In disordered phases the ground state is generically gapped from the rest of the spectrum, and hence even for moderate $\beta$ the thermal density matrix is dominated by the ground state. 

\subsection{Dimerized phase}\label{sec:disordered}
Here we show that measurements in the dimerized phase generate SPT order. We consider the Heisenberg model with $J_2 \gg J_1$; in the limit $1/g = J_1/J_2 \to 0$ the ground state becomes a tensor product of singlets on a nonoverlapping set of bonds, and this dimerized phase extends down to $1/g \approx 0.52$. The ground states with $0 \leq 1/g \lesssim 1/g_c$ are therefore related to the tensor-product ground state at $g=0$ via a finite depth local unitary evolution \cite{bachmann2017adiabatic}. Let us denote the pure ground states of this model by $\ket{\psi_{1/g}}$. Note that since this phase is gapped we have $e^{-\beta H} \sim \ket{\psi_{1/g}}\bra{\psi_{1/g}}$ already for $\beta$ asymptotically larger than $\ln N$.

\begin{figure}[ht!]
    \centering
    \includegraphics[width=0.39\textwidth]{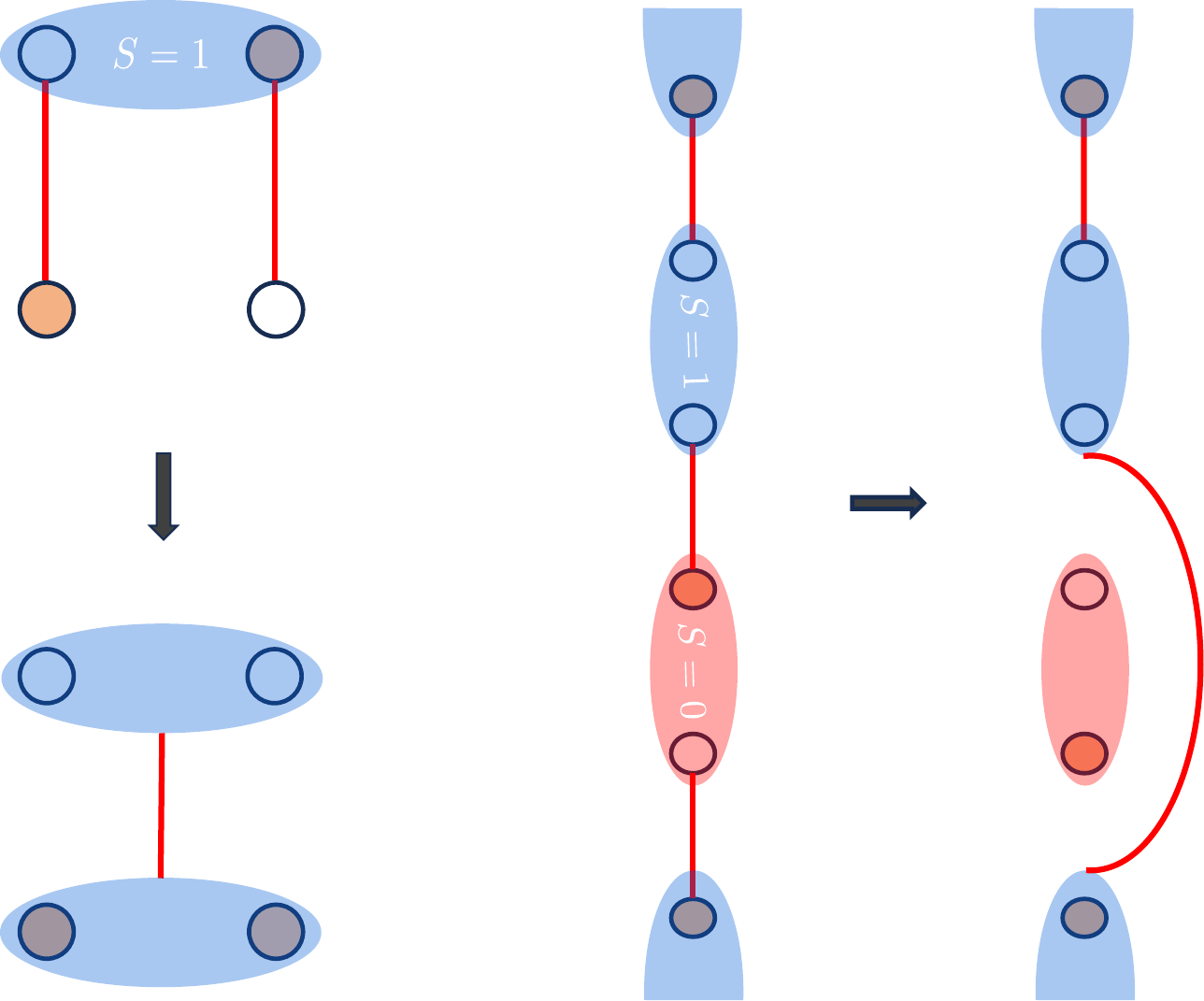} 
    \caption{Left: given a pair of singlets on vertical bonds (red lines), measuring one of the horizontal bonds and finding a triplet outcome leads to the creation of triplet states on both horizontal bonds. Since the measurement operators are symmetric, these two triplets must be entangled together into a total singlet. Right: creating random AKLT chains using \SU{2}-symmetric measurements. Starting from a state consisting of nonoverlapping singlets, and measuring the spin $S^2_{(jk)}$ on pairs of spins which link the original singlets, we either create effective spin-$1$ degrees of freedom (blue ellipses) or longer range singlets (red ellipse and curved red line).}
    \label{fig:illustration_aklt}     
\end{figure}

First consider measuring a state which consists of two singlets, occupying bonds $(0,1)$ and $(2,3)$, respectively. If on measuring $S^2_{(02)}$ we find $s_{02}=0$, the post-measurement state hosts singlets on both $(0,2)$ and $(1,3)$. On the other hand, if we find $s_{02}=1$, then the post-measurement state hosts triplets on both $(0,2)$ and $(1,3)$, with these two triplets entangled together into an overall singlet state. This is illustrated on the left in Fig.~\ref{fig:illustration_aklt}.

Turning now to the full system we will show that, if we measure $S_{(jk)}^2$ on pairs of sites $(jk)$ in the dimerized limit, the states arising from every possible measurement outcome have symmetry-protected topological (SPT) order: Generic measurement outcomes lead to the creation of random arrangements of AKLT chains on the square lattice. A protocol of this kind was recently discussed in Ref.~\cite{jia2024architecture}.

\begin{figure}[!ht]
    \centering
    \includegraphics[width=0.38\textwidth]{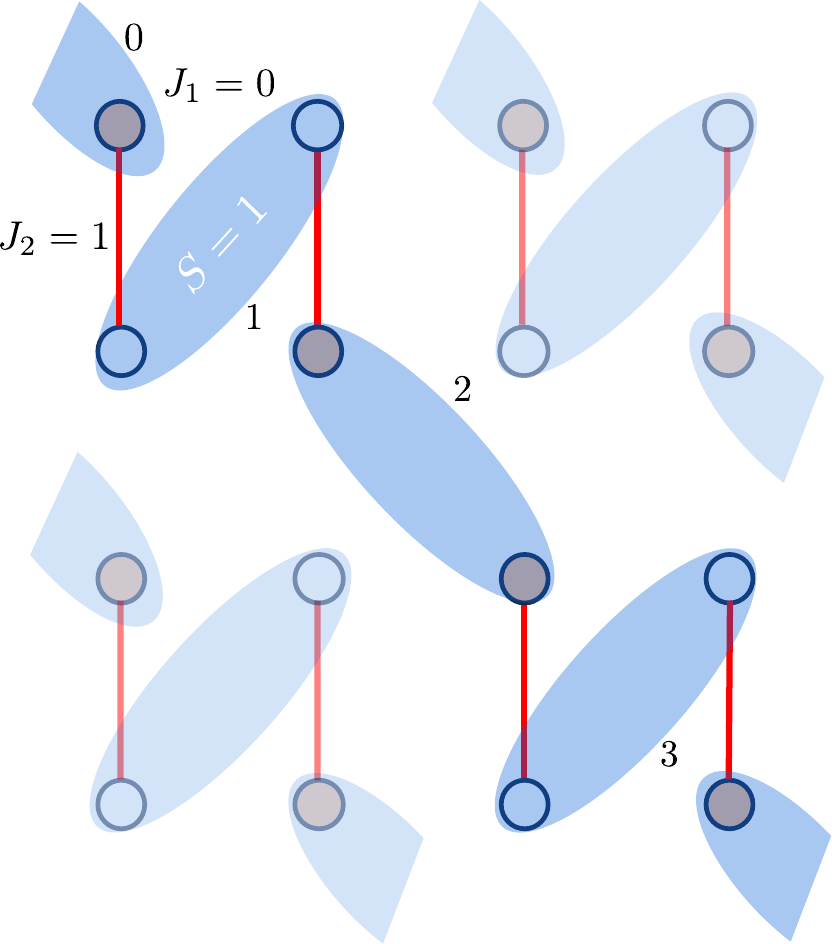}
    \caption{Creating AKLT chains on the square lattice. For general measurement outcomes we create a random arrangement of AKLT chains, and here we illustrate the chains generate by measurements with triplet outcomes on a nonoverlapping set of NNNs. The numbers indicate the lattice indexing relevant to Fig.~\ref{fig:string_order_param}.}
    \label{fig:aklt}
\end{figure}

To see the emergence of AKLT states, imagine starting from a one-dimensional line of nonoverlapping singlets arranged end-to-end on a chain. For example, if our chain has spins at sites $x=0,1,\ldots$ the singlets can be taken to reside on bonds $(x,x+1)$ for $x$ even. Next, consider measuring $S_{jk}^2$ on bonds $(x,x+1)$ with $x$ odd. 

Each measurement has two possible outcomes. On projecting the measured bond $(x,x+1)$ (with $x$ odd) into a singlet $s_{x,x+1}=0$, we also generate a singlet between spins at sites $x-1$ and $x+2$, i.e. local measurement create a nonlocal singlet. On projecting the measured bond $(x,x+1)$ into a triplet, the two spin-$1/2$ degrees of freedom there are instead merged into a single spin-$1$ degree of freedom. We indicate the effects of these measurements on the right in Fig.~\ref{fig:illustration_aklt}.

If we were to measure $s_{x,x+1}=1$ for all odd $x$, our post-measurement state is the AKLT chain. In fact, this is the textbook construction of the AKLT chain as a matrix-product state (MPS) with bond dimension two \cite{cirac2021matrix}: the original singlets of the ground state become the `virtual legs' of the MPS, while the pairs of sites $(x,x+1)$ with $x$ odd become the `physical sites' of the AKLT chain, which has SPT order \cite{gu2009tensor,pollmann2010entanglement}. 

The fact that an AKLT chain is generated for essentially every measurement outcome follows from the `rewiring' property of singlet projections. Suppose that on measuring we find $s_{x,x+1}=1$ for odd $x=x_0,x_1,\ldots$ and that we find $s_{x,x+1}=0$ for all other odd values of $x \neq x_0,x_1,\ldots$. Now the spins at sites $x_k+1$ and $x_{k+1}$ are paired into a singlet. Meanwhile, the spins on the bond $(x_k,x_k+1)$ form an effective spin-$1$ degree of freedom, as do those on the bond $(x_{k+1},x_{k+1}+1)$. We therefore generate an AKLT state consisting of effective spin-$1$ degrees of freedom which reside on the bonds $(x_k,x_k+1)$ of the original chain. The spaces between these bonds become virtual legs in the MPS.

\begin{figure}
    \centering
    \includegraphics{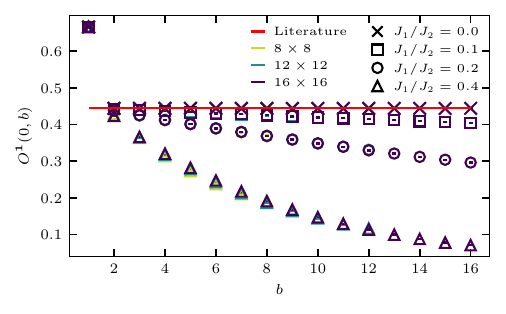}
    \caption{String order parameter $O^{\bm 1}(0,b)$ evaluated after forcing triplets on diagonal bonds, as shown in \Cref{fig:aklt}, shown for various coupling strengths $J_1$. The string order is evaluated in terms of the effective spin-$1$ operators $L^z_b = P^1_{(jk)}(S^z_j + S^z_k)P^1_{(jk)}$ on measured NNN pairs of spins, which we here label by $b$. The index $b=1,2,\ldots$ measures distance along the AKLT chain that is generated by measurement in the case $g=J_1/J_2=0$. We use square $N_x \times N_x$ lattices (see legend) and set $\beta=3N_x$.}
    \label{fig:string_order_param}
\end{figure}

The two-dimensional problem relevant to our model can be understood analogously. If we imagine starting from a set of non-overlapping singlets and choose to measure a set of bonds which, together with the original singlets, form a long `snake' through the lattice [see Fig.~\ref{fig:aklt}], we generate an AKLT state along this snake. Note that the choice of measured pair of sites determines the locations and number of such `AKLT snakes'. This idea can be summarized as
\begin{align}
   P^{\bm{s}}\ket{\psi_0} \propto \bigotimes_{\sigma} \ket{\text{AKLT}^{\bm{s}}}_{\sigma} 
\end{align}
where the index $\sigma$ labels the different snakes, and $\ket{\text{AKLT}^{\bm{s}}}_{\sigma}$ is an AKLT state on the snake $\sigma$, having the locations of its spin-1 degrees of freedom sites fixed by the measurement outcomes $\bm{s}$.

On tuning away from the $1/g=0$ limit, the fate of the SPT order is unclear; this is also the case in one dimension, and so in the following we keep our discussion general. We can write the ground state for $1/g < 1/g_c$ as $\ket{\psi_{1/g}} = U_{1/g} \ket{\psi_0}$, where $U_{1/g}$ is a quasi-local unitary associated with an adiabatic change in the Hamiltonian. The post-measurement state is then $P^{\bm{s}}\ket{\psi_{1/g}} \propto P^{\bm{s}}U_g \ket{\psi_0}$. It is natural to ask how the string order
\begin{align}
    O^{\bm{s}}(a,b) = \Big< L^z_a e^{i\pi \sum_{c=a+1}^{b} L^z_{c}} L^z_{b} \Big>^{\bm{s}}
\end{align}
behaves for $1/g \neq 0$; here $L^z_a = P^{\bm{1}}_{(jk)}(S^z_j + S^z_k)P^{\bm{1}}_{(jk)}$ is a spin-$1$ operator acting on a pair of sites which we here label by $a = (jk)$. For the geometry illustrated in Fig.~\ref{fig:aklt}, the index $a$ labels NNNs, of which feature spin-$1$ degrees of freedom. In Fig.~\ref{fig:string_order_param} we calculate $O^{\bm{s}}(0,b)$ for the post-measurement state corresponding to $\bm{s}=\bm{1}$ on a nonoverlapping set of NNNs. For $1/g=0$ we recover perfect string order as required, and for nonzero $1/g$ within the dimerized phase we find that $O^{\bm{1}}(0,b)$ decays with separation. Here separations are measured along the length of the AKLT chain that would have been created at $1/g=0$.

These numerical calculations have served to demonstrate our algorithm as a probe of correlations in measured quantum states. Next we briefly discuss how the \SU{2} symmetry of our system allows us to calculate mixed-state entanglement measures between pairs of sites using QMC. 
\subsection{Mixed-state entanglement}

A useful consequence of the \SU{2} symmetry of our system is that it allows us to relate standard correlations to entanglement measures. The results of this subsection follow from the fact that any two-qubit density matrix with weak \SU{2} symmetry must have the form
\begin{align}
    \rho^{\bm{s}}_{jk} = \alpha\ket{0,0}\bra{0,0} + \frac{1}{3}(1-\alpha) \sum_m\ket{1,m}\bra{1,m}, \label{eq:rhojk}
\end{align}
with a parameter $\alpha$ that can be inferred from our numerical calculations via the relation $\alpha=\frac{1}{4}-\braket{\bm{S}_j\cdot\bm{S}_k}^{\bm{s}}$. In an abuse of notation here we use the notation $\ket{S,m}$ for the two-qubit states that are eigenstates of $(\bm{S}_j+\bm{S}_k)^2$ and $S^z_j+S^z_k$, where $j$ and $k$ are in general a pair of separated spins. From Eq.~\eqref{eq:rhojk} it can be verified that the von Neumann entanglement entropy of either of the spins $j$ and $k$ is maximal regardless of the value of $\alpha$, i.e. $\rho^{\bm{s}}_{j}=\text{Tr}_k \rho^{\bm{s}}_{jk}=\frac{1}{2}\eye$, where $\text{Tr}_k$ denotes a partial trace over spin $k$. For $\alpha=1$, as arises for $\beta \to \infty$ in the $\bm{s}=\bm{0}$ measurement scheme of Fig.~\ref{fig:proj_corr_heisenberg_ladder}, the spins $j$ and $k$ are maximally entangled with one another. On the other hand, for $\beta=0$, the spins are maximally entangled with the environment, and there is no useful entanglement in the system. 

To quantify the entanglement between $j$ and $k$ when $\rho^{\bm{s}}_{jk}$ is mixed we can consider the negativity \cite{vidal2002computable}. Denoting by $\rho^{T_k}_{jk}$ the partial transpose of $\rho_{jk}$ for spin $k$, the negativity of the mixed state is defined as the sum of the magnitudes of negative eigenvalues of $\rho^{T_k}_{jk}$. Equivalently 
\begin{align}  
    \mathcal{N}_{jk} = \frac{1}{2}\Big( ||\rho^{T_k}_{jk}||_1 - 1\Big).
\end{align} 
The negativity vanishes for separable (i.e. unentangled) mixed states, and therefore can only be nonzero when $j$ and $k$ are entangled. Note, however, that the converse is not true: $j$ and $k$ can be entangled even when $\mathcal{N}_{jk}=0$. For states of the form in Eq.~\eqref{eq:rhojk} it can be verified that $\mathcal{N}_{jk}\neq 0$ provided $\braket{\bm{S}_j \cdot \bm{S}_k} < -\frac{3}{20}$.

Another useful probe of mixed-state entanglement is the coherent information, which is equivalent to the negative conditional entropy. The coherent information between the two spins $j$ and $k$ is defined as
\begin{align}
    I_{j|k} = \mathcal{S}_j - \mathcal{S}_{jk},
\end{align}
where e.g. $\mathcal{S}_j$ is the von Neumann entanglement entropy for $j$. For diagonal (classical) $\rho^s_{jk}$ the right-hand side of this expression reduces to a difference of Shannon entropies, and the Shannon entropy of a system cannot be smaller than that of one of its subsystems. Therefore, in the classical case, $I_{jk}\leq 0$. A positive coherent information therefore indicates entanglement. In fact, $I_{j|k}$ is also a lower bound on the rate of one-way entanglement distillation \cite{devetak2005distillation}. The coherent information can be computed from Eq.~\eqref{eq:rhojk}, and we find that $I_{j|k} > 0$ implies $\braket{\bm{S}_j \cdot \bm{S}_k} \lesssim -0.56$.

Through this section we have studied the symmetry of post-measurement states in both the ordered and disordered phases of the Heisenberg model. Although the ground state is in all cases a total singlet, the presence of the Anderson tower in the ordered phase causes the thermal density matrix to have only a weak \SU{2} symmetry unless $\beta$ is parametrically larger than $N^{d}$, where $d$ is the spatial dimension. In the dimerized model, we showed that the \SU{2} symmetry of the measurements can lead to the creation of SPT order. Finally, we discussed how QMC calculations of standard correlations can be used to determine mixed-state entanglement measures in this setting. 
\section{Enhancing Correlations}\label{sec:triplet}

The effects of measurements on thermal quantum states in $d$ spatial dimensions can be understood qualitatively via a mapping to the statistical mechanics of surfaces \cite{garratt2023measurements}. This connection motivates three related questions, (i) in which classical settings do exotic surface correlations arise, (ii) which of these correlations can be realized in principle as the correlations of post-measurement states, and (iii) in which microscopic models should we expect to find them?

Exotic surface correlations, decaying as an inverse power of a logarithm of separation, have recently been discovered in critical three-dimensional classical \O{n} models \cite{metlitski2022boundary,krishnan2023plane}. The `extraordinary-log' boundary universality class exhibiting these correlations can arise for all $n > 2$ when the couplings on a two-dimensional surface, embedded within a three-dimensional critical bulk, are enhanced relative to those in bulk \cite{krishnan2023plane}. Using the mapping between measurements and the statistical mechanics of surfaces,  Ref.~\cite{lee2023quantum} identified ground states at \O{3} quantum critical points in $d=2$ as starting points from which such correlations could be generated by measurement. However, this possibility was not tested in any microscopic quantum model. Here we study the effects of measuring $S^2_{(jk)}$ for a set of NNN $(jk)$ at the quantum critical point of a Heisenberg AFM, and demonstrate a dramatic measurement-induced enhancement of correlations. 
\subsection{Quantum-classical mapping}

Before discussing our numerical results, we briefly review the quantum-classical mapping as applied to the Heisenberg model, and also the incorporation of measurements into this framework. Recall that, for the thermal state $\rho \sim e^{-\beta H}$, the partition function $Z=\text{Tr}e^{-\beta H}$ of Heisenberg model can be expressed as a spin coherent state path integral \cite{auerbach2012interacting}
\begin{align}
    Z = \int_{\Omega_j(0)=\Omega_j(\beta)} \Big[\prod_j d\bm{\Omega}_j(\tau)\Big] e^{- S[\bm{\Omega}]},
\end{align}
here $\bm{\Omega}_j(\tau)$ are paths of three-component unit vectors parametrized by imaginary time $\tau \in [0,\beta]$, and we have periodic boundary conditions $\bm{\Omega}_j(0)=\bm{\Omega}_j(\beta)$. The action $S[\Omega]=S_1[\Omega]+S_2[\Omega]$ is the sum of two terms,
\begin{align}
    S_1[\Omega] = \sum_{(jk) \in \text{NN}} J_{(jk)} \int_0^{\beta} d\tau\, \bm{\Omega}_j(\tau) \cdot \bm{\Omega}_k(\tau),
\end{align}
and the Berry phase term $S_2[\Omega]$. Neglecting $S_2[\Omega]$ and moving to the continuum limit $\bm{\Omega}_j \to \pm \bm{n}(x_j)$, where the sign depends on the sublattice, we find a nonlinear sigma model (NLSM) with action
\begin{align}
    S_{\text{NLSM}}[\bm{n}] = \int d^2 x d\tau \, \big( [\nabla \bm{n}]^2 + \bm{\dot n}^2\big),
\end{align}
i.e. $Z_{\text{NLSM}} = \int D\bm{n} e^{-S_{\text{NLSM}}[\bm{n}]}$, where the integration is over field configurations with $\bm{n}^2(x)=1$. Here we have omitted an overall prefactor, and we have rescaled $\tau$ so that the spin wave speed is unity. The low-temperature behavior of the two-dimensional Heisenberg antiferromagnet can then be understood, up to corrections arising from the Berry phase, in terms of a classical three-dimensional \O{3} model.

Next we incorporate measurements into this description. Working with weak measurements in the interest of generality, we consider the effect of observing $\bm{s}=\bm{1}$ on a non-overlapping set $B$ of NNN pairs of sites. The measured density matrix for outcome $\bm{s}=\bm{1}$ is $W^{\bm{1}}\rho W^{\bm{1}}$ up to normalization, where $W^{\bm{1}} \sim e^{(\mu/2)\sum_{(jk) \in B} \bm{S}_j\cdot\bm{S}_k}$. Viewing e.g. $e^{(\mu/2)\bm{S}_j\cdot \bm{S}_k}$ as imaginary time evolution for a time $\mu/2$ under $\bm{S}_j \cdot \bm{S}_k$, we find that partition function corresponding to $\text{Tr}\rho^{\bm{s}}$ takes the form
\begin{align}
    Z^{\bm{1}} = \int_{\Omega_j(0)=\Omega_j(\beta+\mu)} \Big[\prod_j d\bm{\Omega}_j(\tau)\Big] e^{- S[\bm{\Omega}]},
\end{align}
where the action is now 
\begin{align}
    S[\Omega]=S_1[\Omega]+S_2[\Omega]-\int_{-\mu/2}^{\mu/2} d\tau \sum_{(jk)\in B} \bm{\Omega}_j\cdot \bm{\Omega}_k,
\end{align}
and the sum is over the set of measured pairs of sites $B$. The coordinate $\tau$ should be understood as running around a circle of circumference $\beta +\mu$, with $\tau=-\mu/2$ equivalent to $\tau=\beta+\mu/2$. The Hamiltonian contribution $S_1[\Omega]$ only involves an integral over the interval $[\mu/2,\mu/2+\beta]$, while the geometric Berry phase term $S_2[\Omega]$ involves an integral over all $\tau \in [0,\beta+\mu]$. 

The projective limit $\mu \to \infty$ here appears to correspond to an infinite range of imaginary times. However, since $\bm{S}_j\cdot\bm{S}_k$ for different $(jk)\in B$ commute with one another, we approach the projective limit already for $\mu$ of order unity, i.e. $e^{(\mu/2)\bm{S}_j\cdot \bm{S}_k}\approx P^1_{(jk)}$ when $\mu \gg 1$. For this reason, we can understand $Z^{\bm{1}}$ as a model that is inhomogeneous in imaginary time and which features a finite-depth `slab' in which the couplings $\bm{\Omega}_j \cdot \bm{\Omega}_k$ between bonds on the same sublattice enhance the N\'eel order. 

Moving to a continuum theory as above, and neglecting the Berry phase, we find an effective model
\begin{align}
    Z^{\bm{1}}_{\text{NLSM}} = \int D\bm{n} \, e^{-S_{\text{NLSM}}[\bm{n}] - \nu\int d^2 x [\nabla \bm{n}]^2}. \label{eq:NLSMmeasure}
\end{align}
Here we have set $\beta \gg \mu$ and replaced the depth-$\mu$ slab by a two-dimensional surface with enhanced coupling $\nu \sim \mu$.

The Berry phase contributions that we have neglected in Eq.~\eqref{eq:NLSMmeasure} will be important to understand e.g. measurement-induced SPT order in the dimerized phase. This can be seen from the fact that the effective model $Z^{\bm{1}}_{\text{NLSM}}$ without the Berry phase term is purely classical. Understanding the effects of the Berry phase on post-measurement states is presumably a rich problem, but it is beyond the scope of the present work. In this section our focus is instead on properties of the post-measurement states that can be understood from the perspective of classical statistical mechanics.

\subsection{Numerical results}

To test the conjecture in Eq.~\eqref{eq:NLSMmeasure} we now compute post-measurement correlations in states with $\bm{s}=\bm{1}$ on NNNs. The basic content of Eq.~\eqref{eq:NLSMmeasure} is that measurements enhance Neel order. We will test for this enhancement in the (ordered) isotropic Heisenberg model in the first instance, and then move to the critical regime.

\begin{figure}
    \centering    \includegraphics{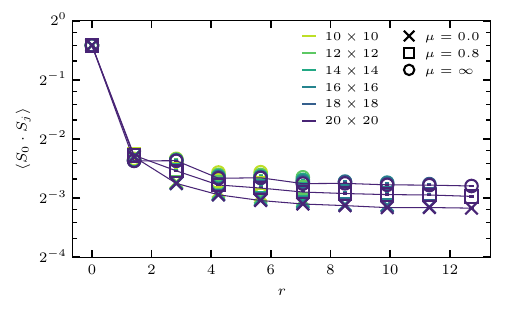}
    \caption{Post-measurement spin correlations $\braket{\bm{S}_0\cdot \bm{S}_j}^{\bm{s}}$ in the isotropic model ($J_1=J_2$) with $\bm{s}=\bm{1}$ on a non-overlapping set of NNNs. The correlation function is calculated along the principal diagonal of a square $N_x \times N_x$ lattice with $N_x=10,\ldots,20$ (see legend), and we set $\beta=3N_x$. The horizontal axis shows the distance $r$ between the sites $0$ and $j$. The long-range order of the unmeasured ($\mu=0$) system is enhanced when the measurement strength $\mu$ is increased; for $\mu=0.8$ and $\mu=\infty$ we use the weak and projective measurement algorithms, respectively.}
    \label{fig:spin corr isotropic heisenberg}
\end{figure}

Our results for the ordered phase are shown in Fig.~\ref{fig:spin corr isotropic heisenberg}. There we find that, as the measurement strength $\mu$ is increased, the post-measurement correlation function $\braket{\bm{S}_j \cdot \bm{S}_k}^{\bm{1}}$ increases. This provides our first evidence in support of the picture in Eq.~\eqref{eq:NLSMmeasure}.

We now turn to the vicinity of the \O{3} quantum critical point, setting $J_2/J_1=g_c$. The results of Ref.~\cite{krishnan2023plane} suggest that, for $\nu > 0$, the model of Eq.~\eqref{eq:NLSMmeasure} develops correlations which decay as
\begin{align}
    \braket{\bm{S}_j \cdot \bm{S}_k}^{\bm{1}} \sim [\ln(r/r_0)]^{-q} \label{eq:extralog}
\end{align}
for distance $r = |\vec{x}_j - \vec{x}_k| \ll N_x,N_y$. Here the exponent $q=0.63(3)$ \cite{sun2023extraordinary}, while $r_0$ is a model-dependent constant. 

\begin{figure}[ht]
\centering\includegraphics{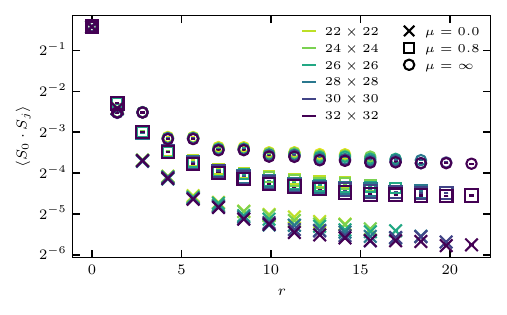}
    \caption{Post-measurement spin correlations in the quantum critical regime, with AFM interaction strengths $J_1=1.91$ and $J_2=1$. As in Fig.~\ref{fig:spin corr isotropic heisenberg} correlations are shown along the diagonal of a square lattice of side $N_x$, and $\beta=3N_x$.}
\label{fig:corr_along_diag_diff_meas_strength}
\end{figure}

\begin{figure}[ht]
    \centering
    \includegraphics{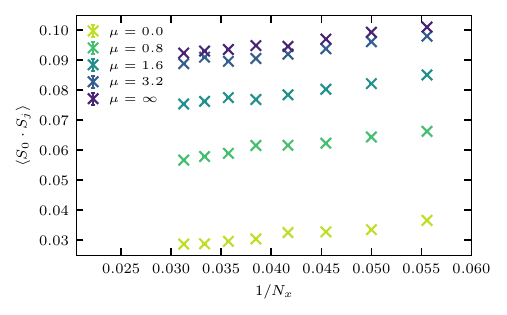}
    \caption{Correlations at fixed distance $r=7\sqrt{2}$ as a function of inverse lattice size $1/N_x$ at the critical coupling studied in Fig.~\ref{fig:corr_along_diag_diff_meas_strength}.
    Increasing the measurement strength $\mu$ causes the correlations to saturate, and we find a similar drift of the correlations with $N_x$ for all $\mu$.}
    \label{fig:corr_fixed_dist_diff_meas_strength}
\end{figure}

In Fig.~\ref{fig:corr_along_diag_diff_meas_strength} we show $\braket{\bm{S}_j \cdot \bm{S}_k}^{\bm{1}}$ for $j$ and $k$ along a diagonal of the lattice for different values of $\mu$ and for various $N_x=N_y$. While at $\mu=0$ correlations are expected to decay as a power law, forcing $\bm{s}=\bm{1}$ with $\mu > 0$ causes correlations to decay much more slowly. This is consistent with the effective model in Eq.~\eqref{eq:NLSMmeasure}, although the precise form of the decay cannot be determined at these system sizes. We probe the finite-size effects in Fig.~\ref{fig:corr_fixed_dist_diff_meas_strength}; there we show the correlations at fixed $r=7\sqrt{2}$ for various $L$, finding an extremely slow drift with increasing $L$. 

\subsection{Ensemble of outcomes}\label{sec:ensemble}
In this section we have provided numerical support for the idea that the effects of measurements can be understood via Eq.~\eqref{eq:NLSMmeasure}. Measuring a NNN pair of sites and finding a triplet outcome appears to enhance the effective coupling on a $(2+0)$-dimensional plane within the $(2+1)$-dimensional theory describing the unmeasured ground state. We have focused on post-selected measurement outcomes in this setting in order to avoid a sign problem. From our numerical results we can nevertheless infer some generic properties of the post-measurement states. 

If we projectively measure $S^2_{(jk)}$ on all NNNs, in general we transform a finite fraction into spin-$1$ degrees of freedom, while the others become local singlets and hence are disentangled from their surroundings. A typical post-measurement state $\rho^{\bm{s}}$ is therefore a random spatial arrangement of spin-$1$ degrees of freedom, and at zero temperature these objects are entangled into a total singlet state. 

We can infer the correlations between the spin-$1$ objects in the post-measurement state as follows. First, denote by e.g. $L^z_{(jk)} = P^{\bm{1}}_{(jk)}(S^z_j+S^z_k)P^{\bm{1}}_{(jk)}$ the spin-$1$ angular momentum operators for the spin-$1$ on bond $(jk)$. These operators commute with $P^{\bm{0}}_{(jk)}$ and $P^{\bm{1}}_{(jk)}$, so we can write
\begin{align}
    \sum_{\bm{s}} p^{\bm{s}}\text{Tr}\big[ \rho^{\bm{s}} \bm{L}_{(jk)} \cdot \bm{L}_{(lm)}] = \text{Tr}\big[ \rho \bm{L}_{(jk)} \cdot \bm{L}_{(lm)}], \label{eq:tripletensemble}
\end{align}
where $\rho$ on the right-hand side is the unmeasured density matrix, $p^{\bm{s}}$ are the Born probabilities, and $\rho^{\bm{s}}=P^{\bm{s}}\rho P^{\bm{s}}/p^{\bm{s}}$. Now let us divide the set of outcomes $\bm{s}$ into three sets, corresponding to (i) $s_{(jk)}=s_{(lm)}=1$, (ii) $s_{(jk)}=s_{(lm)}=0$, (iii) $s_{(jk)} \neq s_{(lm)}$. In cases (ii) and (iii), the contribution of $\bm{s}$ to the left-hand side of Eq.~\eqref{eq:tripletensemble} is zero since $P^{\bm{0}}_{(jk)}\bm{L}_{(jk)} P^{\bm{0}}_{(jk)}=0$; the left-hand side of Eq.~\eqref{eq:tripletensemble} only has contributions from the subset of outcomes for which $s_{(jk)}=s_{(lm)}=1$.

Since only the outcomes with $s_{(jk)}=s_{(lm)}=1$ contribute, spin correlations in the post-measurement states $\rho^{\bm{s}}$ with these outcomes must be enhanced relative to the unmeasured system. For well-separated NNNs $(jk)$ and $(lm)$, a rough approximation for the probability to find $s_{(jk)}=s_{(lm)}=1$ comes from treating the outcomes of the measurements as statistically independent; this probability is  approximately $ \text{Tr}[\rho P^{\bm{1}}_{(jk}]^2 = (3/4+\braket{\bm{S}_j\cdot\bm{S}_k})^2$. Moreover, since the two sites $(jk)$ and $(lm)$ respectively belong to the same sublattices, we can write $\text{Tr}\big[ \rho^{\bm{s}} \bm{L}_{(jk)} \cdot \bm{L}_{(lm)}] \approx 4 \text{Tr}\big[\rho^{\bm{s}} \bm{S}_j \cdot \bm{S}_l\big]$.

Focusing on the critical regime, this relation tells us that for typical measurement outcomes the correlations $\braket{\bm{S}_j \cdot \bm{S}_l}^{\bm{s}}$ between well-separated spins that are not forced into singlets must decay, on average, with a power law prescribed by the bulk \O{3} universality class. However, the amplitude of this decay is enhanced relative to the ground state by the inverse of the probability $\approx \text{Tr}[\rho P^{\bm{1}}_{(jk)}]^{-2}$ to find $s_{(jk)}=s_{(lm)}=1$. The fact that the well-separated spin-$1$ degrees of freedom have power-law correlations on average should be contrasted with the behavior in Eq.~\eqref{eq:extralog}. The stark difference between these two kinds of decay highlights the complicated statistics of correlations within the ensemble of post-measurement states. 

\section{Single-body measurements}\label{sec:ensemble}

In this section we study the effects of measuring local spin operators. Because the unmeasured SSE for the Heisenberg model has no sign problem when working in a fixed site-local basis, measurements in this basis do not introduce a sign problem in the post-measurement SSE. It is important to note that this feature is not specific to the Heisenberg model: If there is a basis in which the unmeasured SSE is sign-free, the effects of measurements in that basis can be studied using a sign-free post-measurement SSE, regardless of the measurement outcomes. Conditional correlations in all post-measurement states can then be evaluated efficiently using our algorithm. 

In the present setting, measurements of local spins break the \SU{2} symmetry of the density matrix and have qualitatively different effects to the symmetric measurements studied in previous sections.  Let us first divide the spins into two sets $A$ and $B$: we will projectively measure $S^z_j$ for spins $j$ in set $B$, and ask how these measurements affect correlations between spins in set $A$. In order to do this we need to (i) sample outcomes of measurements, (ii) calculate post-measurement correlations conditioned on specific sets of measurement outcomes, and (iii) construct a suitable average over these post-measurement correlations.

As described in connection with Eq.~\eqref{eq:sampling}, we sample the outcomes of measurements of $S^z_j$ using the standard (unmeasured) SSE. That is, we sample basis states $\ket{\alpha_A\alpha_B}$ of the full system with probabilities proportional to $\braket{\alpha_A \alpha_B|e^{-\beta H}|\alpha_A\alpha_B}$, and store a random subset of the observed configurations $\ket{\alpha_B}$ of the $B$ spins. Note that here $\alpha_B$ is a set of outcomes $\alpha_j = \pm 1/2$ of $S^z_j$ measurements for all $j \in B$. 

\begin{figure}
    \centering
    \includegraphics[width=0.44\textwidth]{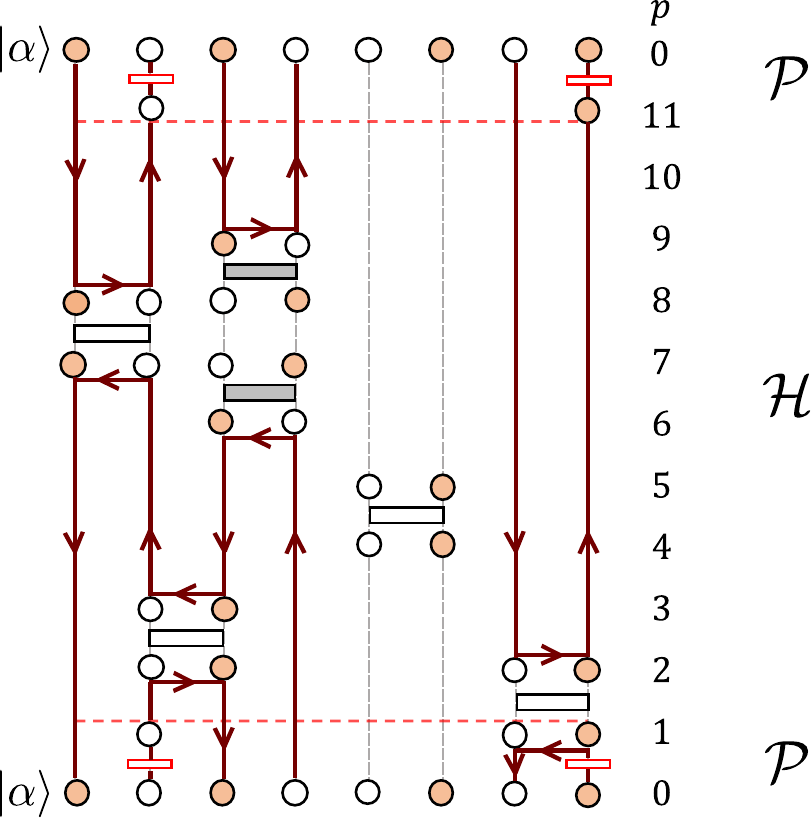}
    \caption{Freezing loop update for single-body $S^z_j$ projectors $\frac{1}{2}(1 \pm 2 S^z_j)$ in the Heisenberg model. A closed loop is initiated at each projector using the deterministic loop \cite{sandvik_computational_2010} update scheme. All operator legs that belong to the constructed loop are frozen—i.e., excluded from further updates—during the off-diagonal update. Frozen legs are indicated in maroon and the arrows show the direction of the loop. Projectors are shown as open red rectangles, and sites with these open red rectangles belong to the set $B$.}
    \label{fig:freeze_loop}
\end{figure}

Each observed set of measurement outcomes $\alpha_j$ defines a set of projection operators $P^{\alpha_j}_j = \frac{1}{2}(1 + 4\alpha_j S^z_j)$, such that the overall projector is
\begin{align}
	P^{\alpha_B} = \bigotimes_{j \in B} P^{\alpha_j}_j.
\end{align}
The partition function which can be used to generate post-measurement correlations is then
\begin{align}
	Z^{\alpha_B} &= \text{Tr}[P^{\alpha_B}e^{-\beta H} P^{\alpha_B}] \\&= \sum_{\alpha_A} \braket{\alpha_A \alpha_B|e^{-\beta H}|\alpha_A \alpha_B}, \notag
\end{align}
and we will only compute correlations between sites $j,k \in A$. Expanding $e^{-\beta H}$ in powers of $\beta$, we can define a post-measurement SSE which samples operator strings as well as configurations $\alpha_A$ of the $A$ sites that are consistent with the fixed configuration $\alpha_B$ of the $B$ sites. To implement this algorithm, during the loop update step we construct closed loops around the $B$ sites at $\tau=0$, thereby ensuring that their states are not updated; we illustrate these loops in Fig.~\ref{fig:freeze_loop}. Ergodicity is preserved over the unmeasured $A$ spins, allowing us to efficiently sample the ensemble of measured states.

Using these methods, we can calculate post-measurement correlations in states $\rho^{\alpha_B} \propto P^{\alpha_B} e^{-\beta H} P^{\alpha_B}$, e.g. $\braket{S^z_j S^z_k}^{\alpha_B} \equiv \text{Tr}[\rho^{\alpha_B}S^z_j S^z_k]$. We now seek a suitable average over correlation functions that can detect the effects of measurements. Denoting the probability to observe $\alpha_B$ by $p^{\alpha_B}$, recall that 
\begin{align}
	\sum_{\alpha_B} p^{\alpha_B} \braket{S^z_j S^z_k}^{\alpha_B} = \braket{S^z_j S^z_k},
\end{align}
i.e. averaging post-measurement correlations $\braket{S^z_j S^z_k}^{\alpha_B}$ according to the Born rule `washes out’ the nonlocal effects of our measurements. A simple probe of the effects of the measurements is the measurement-averaged connected correlation function
\begin{align}
	C_{jk} = \sum_{\alpha_B} p^{\alpha_B} \Big[ \braket{S^z_j S^z_k}^{\alpha_B} - \braket{S^z_j}^{\alpha_B} \braket{S^z_k}^{\alpha_B}\Big].
    \label{eq:meas_avg_correlation}
\end{align}
The second term in the brackets is nonlinear in the post-measurement density matrix, and is generically nonzero. Note that without measurements $\braket{S^z_j}=0$, so this connected correlation function simplifies to $\braket{S^z_j S^z_k}$. 

In practice, we calculate $C_{jk}$ as follows. Initially, we run an unmeasured SSE simulation (Sec.~\ref{sec:intro_SSE}) and sample configurations $\alpha_B$ of the $B$ spins at large intervals along the Markov chain; we denote the number of different samples used by $N_B$. Then for each of the $N_B$ samples, we require estimates for both $\braket{S^z_j S^z_k}^{\alpha_B}$ and $\braket{S^z_j}^{\alpha_B} \braket{S^z_k}^{\alpha_B}$. To evaluate the latter, we run two \emph{statistically independent} post-measurement SSEs: If the errors in our estimates for $\braket{S^z_j}^{\alpha_B}$ and $\braket{S^z_k}^{\alpha_B}$ are of order $\epsilon$, the error in our estimate for $\sum_{\alpha_B} p^{\alpha_B}\braket{S^z_j}^{\alpha_B} \braket{S^z_k}^{\alpha_B}$ is then of order $\epsilon/\sqrt{N_B}$. On the other hand, if our estimates for $\braket{S^z_j}^{\alpha_B}$ and $\braket{S^z_k}^{\alpha_B}$ were correlated, there would be a drift in our estimate for $\sum_{\alpha_B} p^{\alpha_B}\braket{S^z_j}^{\alpha_B} \braket{S^z_k}^{\alpha_B}$ scaling as $\epsilon^2/N_B$; by using two independent Markov chains, we avoid this drift. From the post-measurement SSEs we also obtain estimates for $\braket{S^z_j S^z_k}^{\alpha_B}$, and we average these estimates over observed $\alpha_B$ to estimate $\sum_{\alpha_B} p^{\alpha_B} \braket{S^z_j S^z_k}^{\alpha_B}$. Agreement between the results of this calculation, and the correlations $\braket{S^z_j S^z_k}$ in the unmeasured state, serves as a simple consistency check.

\begin{figure}
    \centering
    \includegraphics[width=0.26\textwidth]{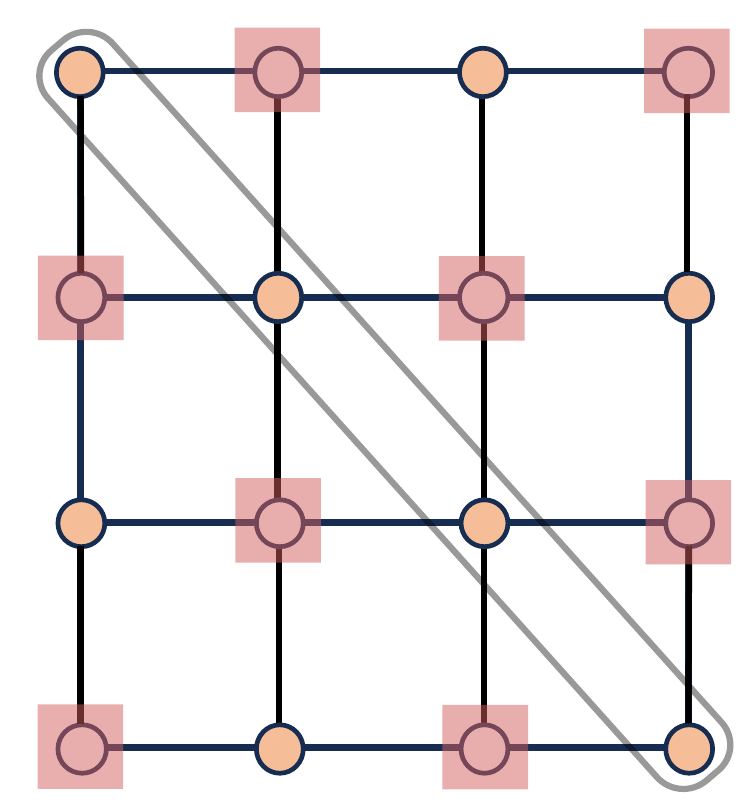}
    \caption{Figure illustrating single site $S^z_j$-measurement setup on square lattice for AFM Heisenberg model. Spins on the $B$ sublattice (indicated with squares) are measured while spins on the $A$ sublattice are not measured. The diagonal enclosed within the rectangular box indicates the direction along which we measure the measurement averaged connected correlation function $C_{0k}$ given by Eq. \eqref{eq:meas_avg_correlation}}
    \label{fig:meas_illus}
\end{figure}

\begin{figure}
    \centering
    \includegraphics{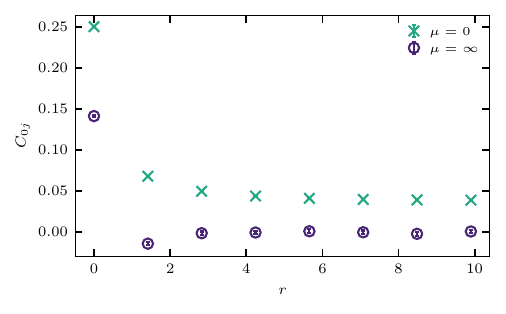}
\caption{
        Connected spin correlations [Eq.~\eqref{eq:meas_avg_correlation}] along the principal diagonal of a square lattice in the quantum critical regime with ($\mu=0$) the $B$ sublattice unmeasured and ($\mu=\infty$) the $B$ sublattice projectively measured in $S^z_j$ basis. Here $N_x=N_y=16$, $\beta = 3N_x = 48$, $J_1=1$, and $J_2=1$. The horizontal axis shows the separation $r$ between sites $0$ and $j$ along the principal diagonal.
    }
    \label{fig:non_linear_avg_ensemble}
\end{figure}

We now calculate $C_{jk}$ for the Heisenberg AFM tuned to the quantum critical regime, with $\alpha_B$ the entire $B$ sublattice, and with spins $j,k \in A$ separated along the principal diagonal of the lattice. The geometry of the set of measured spins $B$ and the principal diagonal along which we measure $C_{jk}$ is shown in Fig. \ref{fig:meas_illus}. To understand what we should expect here, recall that the scaling dimension $\Delta$ of the spin operator at the three-dimensional \O{3} critical point is close to its mean field value of $1/2$, i.e. two point correlations in the unmeasured system decay algebraically with an exponent close to unity. Following Ref.~\cite{garratt2023measurements}, for a quantum critical system in $d$ spatial dimensions, a \emph{weak} measurement of an operator with scaling dimension $\Delta$ constitutes a relevant perturbation if $d > 2\Delta$, which is the case here. Since weak measurements of the spin operator are relevant, it is natural to expect that projective measurements destroy long-range correlations. Therefore, we expect that while the algebraic decay of the algebraic decay of $\braket{S^z_j S^z_k}$ in the unmeasured system will be replaced by a rapid decay of $C_{jk}$ in the post-measurement state. Our numerical results, shown in Fig.~\ref{fig:non_linear_avg_ensemble}, are consistent with this.

In summary, here we have shown that the post-measurement SSE can be used to evaluate correlations in states generated by the measurement of local spin operators in a fixed basis. Here, because there is no sign problem, our algorithm can be used to construct probes of post-measurement correlations~\cite{garratt2023probing,mcginley2024postselection} relevant to experiments in quantum simulators. \\

\section{Discussion}\label{sec:discussion}
In this work we have developed a versatile new QMC technique for the study of measurement-induced quantum phenomena. Our focus was on the effects of measurements on thermal states of many-body systems. At low (or vanishing) temperatures, the strong correlations in these states can amplify the nonlocal effects of local measurements. First, we implemented our technique to probe the effects of \SU{2}-symmetric measurements on Heisenberg AFMs on the square lattice. For this model we explored the role of the Anderson tower of states, the creation of SPT order via measurement, and the possibility of creating quantum states whose correlations realize exotic surface universality classes. Although these measurements typically generate a QMC sign problem when working a site-local basis, there is no sign problem when performing single spin measurements in a fixed basis. Using such a scheme, and studying averaged properties of the resulting ensemble of post-measurement states, we showed numerically that \O{3} quantum critical correlations are fragile to measurements of the order parameter.

The post-measurement SSE allows for the efficient evaluation of expectation values in measured quantum states when there is no QMC sign problem. 
At a high level, the idea is to expand the measured density matrix as a sum over large numbers of `strings' (products) of local operators, and to express expectation values in the measured density matrix as averages over contributing operator strings. For large classes of density matrices, this decomposition can be performed in such a way that the averages can be evaluated over the steps of a Markov chain, i.e. via Monte Carlo sampling of operator strings. The existence of a decomposition with this property amounts to the absence of a `sign problem'.

When applied to thermal density matrices, this is the SSE formulation of QMC \cite{sandvik_computational_2010}. There the density matrix is expressed as a sum over powers of the Hamiltonian, itself a sum of local operators. Here, the measured density matrix is itself a product of (i) the initial (in our case thermal) density matrix and (ii) the Kraus operators which, through their action on the left and on the right of the density matrix, describe a specific set of (projective or weak) measurement outcomes. Each of (i) and the two copies of (ii) are expanded as sums over operator strings, and the products of these three sets of strings define our expansion of the measured density matrix. As for the thermal SSE, the basic technical problem is to identify a choice of local operators in such a way that the operator strings contributing to the density matrix can be sampled.

For \SU{2}-symmetric local measurements of the total spin $S^2_{(jk)}$ on pairs of sites, and when working in a fixed site-local basis, we identified a sign problem for bipartite Heisenberg AFMs. Without measurements, there is no sign problem in this model. The sign problem generated here by measurements has a similar form to that encountered in standard thermal QMC for frustrated antiferromagnets, and these have remained unresolved for decades \cite{sandvik_computational_2010}. However, an important difference here is that the contributions to the SSE which generate the sign problem only appear during a finite interval of imaginary time, whereas in the thermal SSE for a frustrated AFM they appear everywhere in imaginary time. It would be interesting to understand whether this difference allows for a resolution of the measurement-induced sign problems encountered here. We note, however, that the sign problem does not appear when measuring in a fixed site-local basis, e.g. when measuring $S^z_j$ on many sites $j$.

In systems without a sign problem, our method allows for new experiments studying measurement-induced phenomena in quantum simulators. The post-selection problem, arising from the fact that the Born probabilities of extensive sets of measurement outcomes are exponentially small in system size, means that individual post-measurement states are unlikely to be observed more than once (unless the experiment is repeated exponentially many times). Cross-correlations between the results of the experiment and numerical estimates for post-measurement expectation values (or density matrices) can nevertheless bound properties of the ensemble of post-measurement states \cite{garratt2023probing,mcginley2024postselection}. Efficient numerical techniques for estimating these expectation values, beyond one spatial dimension and for low-energy states of generic interacting systems, were previously absent. Low-energy states of many-body quantum systems have nevertheless been prepared in a variety experiments with access to local measurements. The post-measurement SSE algorithm can immediately be applied to observe measurement-induced collective phenomena in those settings.

The correlations which can be probed in experiment are nonlinear functions of post-measurement density matrices, e.g. averages of $\braket{S^z_j}^{\bm{s}}\braket{S^z_k}^{\bm{s}}$ and the entanglement entropy $-\text{Tr}[\rho^{\bm{s}}\log \rho^{\bm{s}}]$ over observed outcomes $\bm{s}$. Nonlinear averages of this kind can have rather exotic behavior, with stark differences between different post-measurement states as discussed throughout this work. Notably, using techniques from conformal field theory (CFT), Ref.~\cite{patil2024highly} recently identified one-dimensional critical ground states where these nonlinear correlation functions exhibit multifractal scaling. Exploring the statistical properties of measurement-averaged correlation functions using our algorithms is promising direction for future work. 

Some of the most striking effects of measurements on ground states have been identified in calculations of post-measurement entanglement \cite{lin2023probingsign,weinstein2023nonlocality,murciano2023measurement,sun2023new,yang2023entanglement,cheng2024universal,hoshino2024entanglement,patil2024highly}. For example, when measuring a finite fraction of the degrees of freedom in a one-dimensional quantum critical state \cite{weinstein2023nonlocality,murciano2023measurement,sun2023new,yang2023entanglement,patil2024highly} it has been shown that measurements which act as marginal boundary perturbations can alter the effective central charge, which characterizes the logarithmic increase of entanglement entropy with subsystem size, while measurements acting as irrelevant boundary perturbations do not appear to change this quantity. On the other hand, relevant perturbations have been shown to drive states to area-law entanglement. By generalizing QMC calculations of entanglement entropies \cite{melko2010finite,hastings2010measuring,humeniuk2012quantum} to the post-measurement setting, it is natural to ask about the effects of measurements on entanglement in higher spatial dimensions. 

The data and code used to generate the plots in this paper are publicly available at Zenodo \cite{baweja2024data}.

\begin{acknowledgments}
KB and DJL were supported by the Deutsche Forschungsgemeinschaft through the cluster of excellence ML4Q (EXC 2004, project-id 390534769), the QuantERA II Programme that has received funding from the European Union's Horizon 2020 research innovation programme (GA 101017733), the Deutsche Forschungsgemeinschaft through the project DQUANT (project-id 499347025), and the Deutsche Forschungsgemeinschaft through CRC 1639 NuMeriQS (project-id 511713970) and CRC TR185 OSCAR (project-id 277625399). SJG was supported by the Gordon and Betty Moore Foundation, and by the U.S. Department of Energy, Office of Science, Office of High Energy Physics, under QuantISED Award DE-SC0019380.

\end{acknowledgments}

\appendix

\section{Loop updates}\label{sec:loop_appendix}
Here we discuss the loop updates used for the QMC calculations throughout this work. These updates are crucial for achieving ergodicity and also for maintaining computational efficiency when sampling operator string configurations composed of both diagonal and off-diagonal Hamiltonian terms. 

Recall that, in the SSE, the full set of operator strings acting on a basis state must return that state to itself, i.e. there are periodic boundary conditions in imaginary time. Inserting off-diagonal operators into the string while respecting these periodic boundary conditions is, at least naively, straightforward using local updates, such as pairwise insertions and deletions of these operators. Unfortunately, such updates typically fail to sample the full configuration space. This is because local updates cannot alter global topological features of the operator sequence, such as loop winding numbers. 

Loop updates address this challenge by constructing and flipping non-local clusters of spins or operator segments. These updates satisfy detailed balance while enabling transitions between topological sectors, thereby ensuring that all relevant operator string configurations are accessible. In practice, a full loop update sweep visits each operator, on average, once, resulting in a computational cost that scales linearly with both the number of lattice sites $N$ and the inverse temperature $\beta$, that is, $\mathcal{O}(\beta N)$.

In the parameter regimes explored in this work, we observe no signs of critical slowing down. Loop updates remain effective even in the presence of long-range correlations and near-critical behavior. To ensure convergence to equilibrium, we follow standard QMC procedures by monitoring observables as a function of $\beta$. We find that choosing $\beta \propto 3N$ is sufficient to obtain accurate results for all observables of interest.

\section{Inhomogeneous imaginary time} \label{sec:weak_meas_appendix}

In this Appendix we formulate a SSE algorithm for density matrices generated by imaginary-time evolution under multiple distinct operators, and show how it can be used to analyze the effects of weak measurements. In practice, this is the numerical implementation which we use for finite $\mu$ calculations; another formulation was discussed in Sec.\ref{sec:weak}

As previously discussed the key idea of post-measurement SSE is to write the post-measurement partition function as a sum over operator strings. Here, the partition function takes the form
\begin{equation}
    Z^{\bm{s}}_{\mu}=\mathrm{Tr}\Big(e^{-\mu M^{\bm{s}}}e^{-\beta H}\Big),
    \label{eqn:meas_part_func}
\end{equation}
where $M^{\bm{s}} = \sum_{(jk) \in B} M^{s_{(jk)}}_{(jk)}$ and
\begin{align}
    M^{s_{(jk)}}_{(jk)} = (1-2s_{(jk)}) \bm{S}_j\cdot\bm{S}_k.
\end{align}
The goal is now to express $Z^{\bm{s}}_{\mu}$ in a form suitable for SSE implementation. To do so, we start by Taylor expanding both the exponentials present in Eq.~\ref{eqn:meas_part_func}. This gives
\begin{equation}
    Z^{\bm{s}}_{\mu}=\sum_{m,n=0}^{\infty}\frac{(-\mu)^m (-\beta)^n }{n!m!}\mathrm{Tr}\Big((M^{\bm{s}})^mH^n\Big),
    \label{eqn:part_function_w_meas}
\end{equation}
where we have two expansion orders $n$ and $m$ corresponding to $H$ and $M^{\bm{s}}$ respectively. This implies that in the post-measurement partition function described in Eq.~\ref{eqn:part_function_w_meas} the measurement strength $\mu$ has a role analogous to $\beta$ i.e. it enters the calculation as a kind of `imaginary time'. The length of the measurement string is therefore variable. For the weak measurement case the maximum required expansion orders $n$ and $m$ are controlled by $\beta$ and $\mu$. 

We now split the operators $M^{s}_{(jk)}$ into terms $O^s_{d,(jk)}$ which do not cause branching in the computational basis. The operators $O^0_{d,(jk)}$ and $O^1_{d,(jk)}$ are as defined in Eqs.~\eqref{eq:singletdecomposition} and \eqref{eq:tripletdecomposition}, respectively. For example, in the cases where all outcomes are singlets ($\bm{s}=\bm{0}$) or all are triplets, ($\bm{s}=\bm{1}$), we have
\begin{align}
    \begin{split}
      M^{\bm{0}}&=\sum_{d=1,2}(-1)^d\sum_{(jk)\in B}\frac{1}{2}O_{d,(jk)}^{0},\\
      M^{\bm{1}}&=-\sum_{d=1,2}\sum_{(jk)\in B}\frac{1}{2}O_{d,(jk)}^{1}.
      \end{split}
    \label{eqn:antiferro_meas_ham}
\end{align}
where we have omitted additive constants. Inserting decompositions of this kind into Eq.~\ref{eqn:part_function_w_meas} we find a sum over operator strings $\mathcal{M}_{\bm d}^{\bm{s}}\mathcal{H}_{\bm a}$, where the form of $\mathcal{H}_{\bm a}$ is given by Eq.~\ref{eqn:opstring_H} while $\mathcal{M}_{\bm d}^{s}$ is
\begin{align}
\mathcal{M}_{\bm{d}}^{\bm{s}}=\prod_{p=L^{H}_{\bm{d}}}^{L-1}O_{d_p,(jk)_p}^{s_{(jk)}}.
    \label{eqn:opstring_M}
\end{align}
As before, the operator type can be $d_p=0,1,2$ where $d_p=0$ is used to refer to identity operator. As usual, the index $\bm{d}$ is an ordered list of operator types $d_p$ as well the pairs of sites on which these operators act. We again work with fixed length operator strings, where now total number of operators $L=L_{\bm d}^{H}+L_{\bm d}^{M}$, while $L^{M}=\sum_{d=0}^{2}L_{d}^{M}$ and $L^{H}=\sum_{d=0}^{2}L_{ d}^{H}$ the fixed total lengths of $\mathcal{M}_{\bm{d}}^{\bm{s}}$ and $\mathcal{H}_{\bm{d}}$

We can assign statistical weights to the operator strings as follows. First, the weight $\mathcal{C}^H_{\bm{a}}$ for $\mathcal{H}_{\bm{a}}$ is given by Eq.~\ref{eqn:stat_weight_H}. The statistical weight $\mathcal{C}^{\bm{s}}_{\bm{d},\mu}$ of $\mathcal{M}_{\bm{d}}^{\bm s}$ now takes a similar form to that of $\mathcal{H}_{\bm{d}}$, and this can be computed directly from Eq.~\eqref{eqn:antiferro_meas_ham} above. For example, with $\bm{s}=\bm{0}$ we have
\begin{align}
    \mathcal{C}^{\bm{0}}_{\bm{d},\mu} = (\mu/2)^{L^M_1+L^M_2} (-1)^{L_2^M} \frac{L^M_0!}{L^M!},
    \label{eqn:weight_weak_meas}
\end{align}
where $L^M_d$ is the number of type $d$ operators in the string $\mathcal{M}^{\bm{s}}_{\bm{d}}$ indicated on the left. Note also that the matrix element $\braket{\alpha|\mathcal{M}^{\bm{s}}_{\bm d}\mathcal{H}_{\bm a}|\alpha}$ remains a boolean variable, i.e. it is unity for allowed configurations and zero otherwise. The post-measurement partition functions for singlet and triplet outcomes differ in the sign factors accompanying type $d=2$ operators, as well as in the configurations which vanish when acted on by $d=1$ operators.

\begin{figure}
    \includegraphics[width=0.44\textwidth]{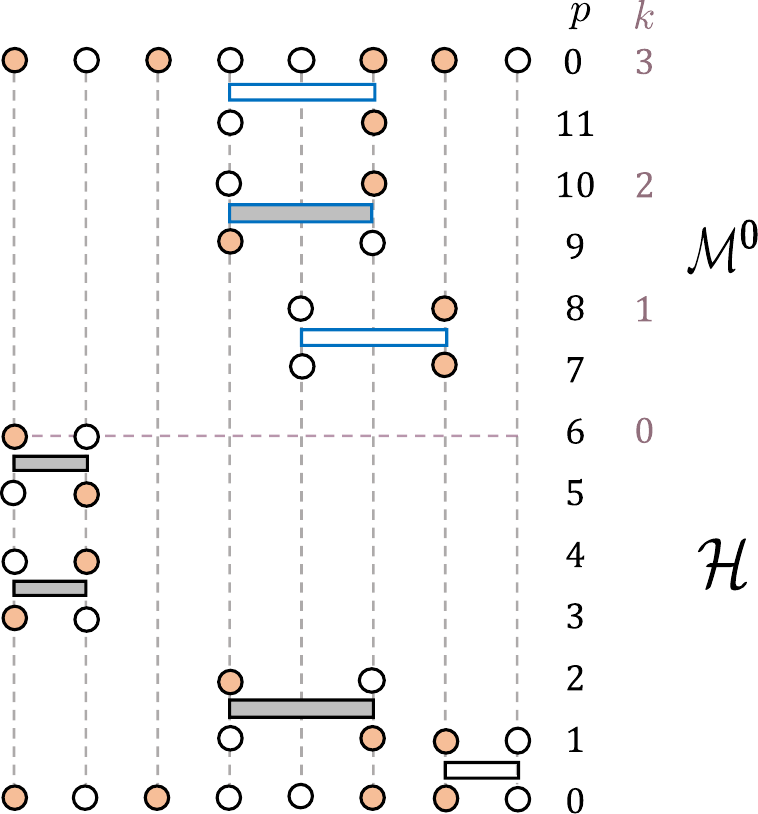}
    \caption{Example SSE configuration within the inhomogeneous imaginary time scheme for the $2 \times 6$ lattice shown in \Cref{fig:benchmarking_models}. Here $\bm{s}=\bm{0}$ and the operator string consists of $\mathcal{H}_{\bm a}$ and $\mathcal{M}^{\bm{0}}_{\bm d}$. The propagated state $\ket{\alpha(6)}$, indicated by purple dashed line, separates the $\mathcal{H}_{\bm d}$ and $\mathcal{M}^{\bm{0}}_{\bm d}$ operator strings.}
    \label{fig:SSE_meas_config}
\end{figure}

The full partition function in the fixed length scheme is then
\begin{align}
       Z^{\bm s}_{\mu}=&\sum_{\substack{\alpha,\bm a,\bm d}} \mathcal{C}^{H}_{\bm{a}}\mathcal{C}^{\bm{s}}_{\bm{d},\mu}\langle \alpha| \mathcal{M}^{\bm{s}}_{\bm{d}} \mathcal{H}_{\bm{a}}|\alpha\rangle.
       \label{eqn:afm_meas_fixed_length_part_func}
\end{align} 
As usual, the periodicity in imaginary time is satisfied by $\mathcal{H}_{\bm a}$ and $\mathcal{M}^{\bm{s}}_{\bm d}$ operator strings together, i.e. $\ket{\alpha(0)}=\ket{\alpha(L-1)}=\ket{\alpha}$, and we illustrate an example configuration in  \Cref{fig:SSE_meas_config}. The diagonal and off-diagonal updates used to sample operator strings contributing to this partition function have the same form as described in Sec.~\ref{sec:implementation}.

\section{Post-measurement expectation values within inhomogeneous imaginary time scheme}

Following weak measurements of the kind described in Appendix~\ref{sec:weak_meas_appendix}, the expectation value for an observable $O$ is given by
\begin{equation}
    \langle O \rangle^{\bm s}_{\mu}=\frac{\mathrm{Tr}\left( e^{-\beta H}e^{-\mu M^{\bm{s}}/2} O e^{-\mu M^{\bm{s}}/2}\right)}{\mathrm{Tr}\left( e^{-\beta H}e^{-\mu M^{\bm{s}}}\right)}.
    \label{eqn:expec_post_meas}
\end{equation}
To study the post-measurement state and compute its expectation values, a modified approach is necessary. While it may seem that expectation values must now be calculated at a single imaginary-time slice, doing so would substantially increase computational time. To mitigate this, we can reduce the time spent by increasing the number of Monte Carlo measurements for a given configuration.

We achieve this by generalizing the method outlined in Ref.~ \cite{luitz2014}, which provides a detailed approach for calculating improved estimators for entanglement and Rényi entropy, leveraging translational symmetries in both real and imaginary time.

From Eq.~\eqref{eqn:expec_post_meas} it is clear that we need to simulate two $\exp(-\mu M^{\bm s}/2)$ strings to calculate the expectation value. However, in \cref{eqn:afm_meas_fixed_length_part_func}, only one string $\exp(-\mu M^{\bm s})$ appears. To correctly compute the expectation value we introduce the notion of compressed imaginary time denoted by $k$. The integer index $k=0,1,\ldots$ counts non-identity operators in the $\mathcal{M}^{\bm{s}}_{\bm{d}}$ string; see Fig.~\ref{fig:SSE_meas_config}) for an illustration. Expectation values are then estimated using
\begin{equation}
    \langle O\rangle^{\bm s}_{\mu}=\frac{1}{2^m}\sum_{k=0}^{m}{m\choose k} O_{k}.
    \label{eqn:obs_expec_val}
\end{equation}
Here, $m$ represents the expansion order of the measurement string with a total length $L^{M}_{\bm d}$, while $O_{k}$ is the expectation value in the propagated state corresponding to `compressed imaginary-time' index $k$. 

Equation~\eqref{eqn:obs_expec_val} is derived by matching the expansion orders of two copies of $\exp(-\frac{\mu}{2}M^{\bm s})$ with $\exp(-\mu M^{\bm s})$. The constraint is that the expansion orders of the two $\mu/2$ strings, which we denote $m_1$ and $m_2$, must sum to the expansion order of the full measurement string: $m = m_1 + m_2$. We can view $m_1$ as being equal to the compressed imaginary time $k$, while $m_2=m-k$.

\section{Additional numerical calculations}
\begin{figure}
\includegraphics{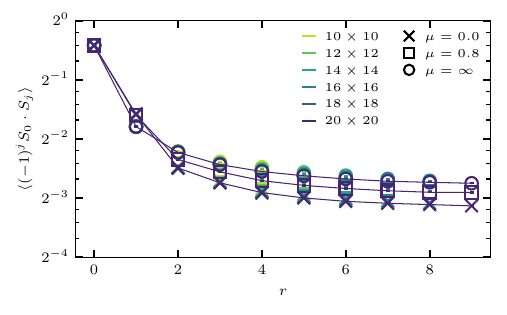}
\includegraphics{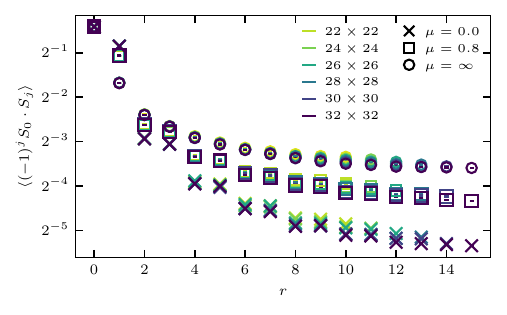}
 \caption{Spin correlations along the vertical of $N_x\times N_x$ periodic lattice for different measurement strengths with outcomes $\bm{s}=\bm{1}$ on NNNs [see Fig.~\ref{fig:aklt}]. (Upper) Isotropic model. (Lower) Critical regime of columnar dimer model.}
 \label{fig:corr_columnar_heisenberg_x_y}
\end{figure}
Here we provide additional calculations of post-measurement correlation functions following $\bm{s}=\bm{1}$ outcomes on NNNs. In Figs.~\ref{fig:spin corr isotropic heisenberg} and ~\ref{fig:corr_along_diag_diff_meas_strength} we calculated post-measurement correlations $\braket{\bm{S}_0\cdot \bm{S}_j}^{\bm{1}}$ along the diagonal of a square lattice for the isotropic model as well the partially dimerized model in its quantum critical regime. Here, we show correlations along the vertical. The results are shown in Fig.~\ref{fig:corr_columnar_heisenberg_x_y}. As in Figs.~\ref{fig:spin corr isotropic heisenberg} and ~\ref{fig:corr_along_diag_diff_meas_strength}, we find that forcing triplet outcomes $\bm{s}=\bm{1}$ on NNNs enhances the antiferromagnetic correlations, and that this enhancement is more significant in the critical regime. 

\end{document}